\documentclass[onecolumn]{aastex631}
\usepackage{url}
\usepackage{xcolor}
\usepackage{CJK}
\usepackage{amsmath,amssymb}

\usepackage{comment}

\shorttitle{A Systematic Study of Millihertz Quasiperiodic Oscillations in GS 1826$-$238}
\shortauthors{Xiao et al}

\graphicspath{{./}{figures/}}

\begin{document}

\title{A Systematic Study of Millihertz Quasiperiodic Oscillations in GS 1826-238}

\author[0009-0004-1288-4912]{Hua Xiao}
\affiliation{School of Physics and Astronomy, Sun Yat-sen University, Zhuhai, 519082, People's Republic of China}
\affiliation{Department of Physics and Astronomy, FI-20014 University of Turku,  Finland}

\author[0000-0001-9599-7285]{Long Ji\textsuperscript{*}}
\email{jilong@mail.sysu.edu.cn}
\affiliation{School of Physics and Astronomy, Sun Yat-sen University, Zhuhai, 519082, People's Republic of China}
\affiliation{CSST Science Center for the Guangdong-Hong Kong-Macau Greater Bay Area, DaXue Road 2, 519082, Zhuhai, People's Republic of China}

\author[0000-0002-9679-0793]{Sergey Tsygankov}
\affiliation{Department of Physics and Astronomy, FI-20014 University of Turku,  Finland}

\author{Yupeng Chen}
\affiliation{Key Laboratory of Particle Astrophysics, Institute of High Energy Physics, Chinese Academy of Sciences, Beijing 100049, People's Republic of China}

\author{Shu Zhang}
\affiliation{Key Laboratory of Particle Astrophysics, Institute of High Energy Physics, Chinese Academy of Sciences, Beijing 100049, People's Republic of China}

\author[0000-0003-2310-8105]{Zhaosheng Li}
\affiliation{Key Laboratory of Stars and Interstellar Medium, Xiangtan University, Xiangtan 411105, Hunan, People's Republic of China}

\begin{abstract}
We performed a systematic investigation of millihertz quasiperiodic oscillations (mHz QPOs) in the low-mass X-ray binary GS 1826$-$238 observed with {\it NICER} and {\it Insight}-HXMT. 
We discovered 37 time intervals exhibiting mHz QPOs out of 106 Good Time Interval (GTI) samples in the frequency range of 3$-$17 mHz at a significance level of $>99.99\%$. 
The source remains in a soft state in our study. 
No significant differences are found between the samples with and without mHz QPOs according to positions in the color-color and hardness-intensity diagrams.
These QPOs were discovered at an accretion rate of $\sim 0.1 \dot{M}_{\rm Edd}$, similar to other sources. 
The broadband spectrum of GS 1826$-$238 can be modeled as a combination of a multicolor blackbody from the accretion disk and a Comptonization with seed photons emitted
from the neutron star (NS) surface. 
The flux modulations of mHz QPOs are related to variations of the temperature of Comptonization seed photons, consistent with the marginally stable burning theory.
\end{abstract}


\section{Introduction} \label{sec:intro}

Low mass X-ray binaries (LMXBs) consist of a compact object, either a NS or a black hole (BH), accreting material from a companion star. 
If the compact object is an NS, the accreted material is accumulated on its surface, potentially triggering thermonuclear X-ray bursts (also known as Type I bursts) under certain conditions \citep[for a review][]{Galloway2021}. 
Depending on the accretion rate, a Type I burst occurs when the accretion rate remains below a critical threshold \citep{Fujimoto1981, Bildsten1998}. At a higher accretion rate, stable thermonuclear burning can take place, preventing the occurrence of Type I bursts.
Near the critical threshold, another type of variability, known as millihertz quasiperiodic oscillations (mHz QPOs), is observed in NS-LMXBs.
They were first discovered in 4U 1608$-$52, 4U 1636$-$53 and Aql X-1 at a frequency of 7–9 mHz \citep{Revnivtsev2001}, and then were reported in other sources, such as 4U 1323$-$619 \citep{Strohmayer2011}, IGR J00291+5934 \citep{Ferrigno2017}, GS 1826$-$238 \citep{Strohmayer2018}, EXO 0748$-$67 \citep{Mancuso2019}, 1RXS J180408.9$-$342058 \citep{Tse2021} and 4U 1730$–$22 \citep{Mancuso2023}.
They are thought to originate from a special mode of nuclear burning on the NS surface \citep{Revnivtsev2001}.  
Through one-zone and multizone numerical simulations, \citet{Heger2007} demonstrated  that marginally stable nuclear burning on the NS surface could give rise to mHz QPOs, and this hypothesis is supported by the fact that mHz QPOs are only detected in a specific range of X-ray luminosities \citep[see Table 1 in][]{Tse2021}. 
However, theoretical calculations suggest that the marginally stable burning should appear near the Eddington accretion rate ($\dot{M}_{\rm Edd}$) \citep{Bildsten1998}. 
This contrasts with observations, which indicate that mHz QPOs are more commonly found at accretion rates around 0.1$\dot{M}_{\rm Edd}$ {\citep[e.g.,][]{Revnivtsev2001,Yu2002,Altamirano2008,Strohmayer2018,Mancuso2019,Mancuso2023,Tse2021}.} 
This discrepancy can be alleviated if the marginally stable burning only occurs
on a part of the NS surface \citep{Heger2007} or taking the effect of turbulent rotational mixing of the accreted material and higher heat flux from deeper layers into account \citep{Keek2009}.
The nuclear burning mechanism is further supported by the anticorrelation between the frequency of kHz QPOs and the X-ray flux modulated by mHz QPOs, because during each mHz QPO the radiation enhancement from the NS surface provides a stronger radiation pressure pushing the inner disk outwards, resulting in a decrease in the frequency of kHz QPOs \citep{Yu2002}.
\citet{Stiele2016} performed a phase-resolved spectral analysis of mHz QPOs in 4U 1636$-$53 and found that variations in the blackbody area on the NS surface lead to mHz QPOs. However, \citet{Strohmayer2018} pointed out that the oscillations in GS 1826$-$238 are consistent with the modulation of blackbody temperature, which supports the model proposed by \citet{Heger2007}. 

The frequency drifting of mHz QPOs was found in 4U 1636$-$53, EXO 0748-676 and Aql X-1 \citep{Altamirano2008,Lyu2015,Mancuso2019,Mancuso2021}.
In 4U 1636$–$53, an X-ray burst would appear if the mHz QPO frequency drops below a certain value. 
After the burst, the mHz QPO is absent for a while and reappears before the next burst, indicating a connection between the mHz QPOs and the unstable nuclear burning on the NS surface.
Such a drifting is accompanied by the movement of the position in the color–color diagram (CCD). \citet{Keek2009} found that the frequency is correlated with the heat flux from the crust, suggesting that the frequency drift could be due to the cooling of the deeper layers where nuclear burning occurs.
GS 1826$-$238 is a typical NS-LMXB, located at a distance of 5.7 kpc \citep{Chenevez2016}. It is also known as {\textquotedblleft}clocked burster{\textquotedblright} due to the presence of extremely regular X-ray bursts \citep{Ubertini1999}.
%
GS 1826$-$238 exhibited a hard spectral state for most of the period before 2016 and was classified as an atoll source. 
A short-term soft spectral state was reported in June 2014 based on the observations of {\it Swift} and {\it NuSTAR} \citep{Chenevez2016}, accompanied with changes of burst behaviors (i.e., periodicity, recurrence time, and profile).
After 2016, this source transfers into intermediate and soft states \citep{Ji2018,Yun2023}.
\citet{Strohmayer2018} reported the detection of mHz QPO at 9 mHz in this source for the first time and its mHz QPO properties support the marginally stable burning model.

In this paper, we present an extensive investigation of mHz QPOs in GS 1826$-$238 observed with {\it NICER} and {\it Insight}-HXMT. This paper is structured as follows. In Section \ref{sec2} we describe the observations and data reduction used in this study and in Section \ref{sec3} we present the main results. Finally, we discuss the observational findings and their implications in Section \ref{sec4}.

\section{Observations and Data Reduction \label{sec2}}
\subsection{{\it NICER}}
\begin{figure}
	\includegraphics[width=0.8\columnwidth]{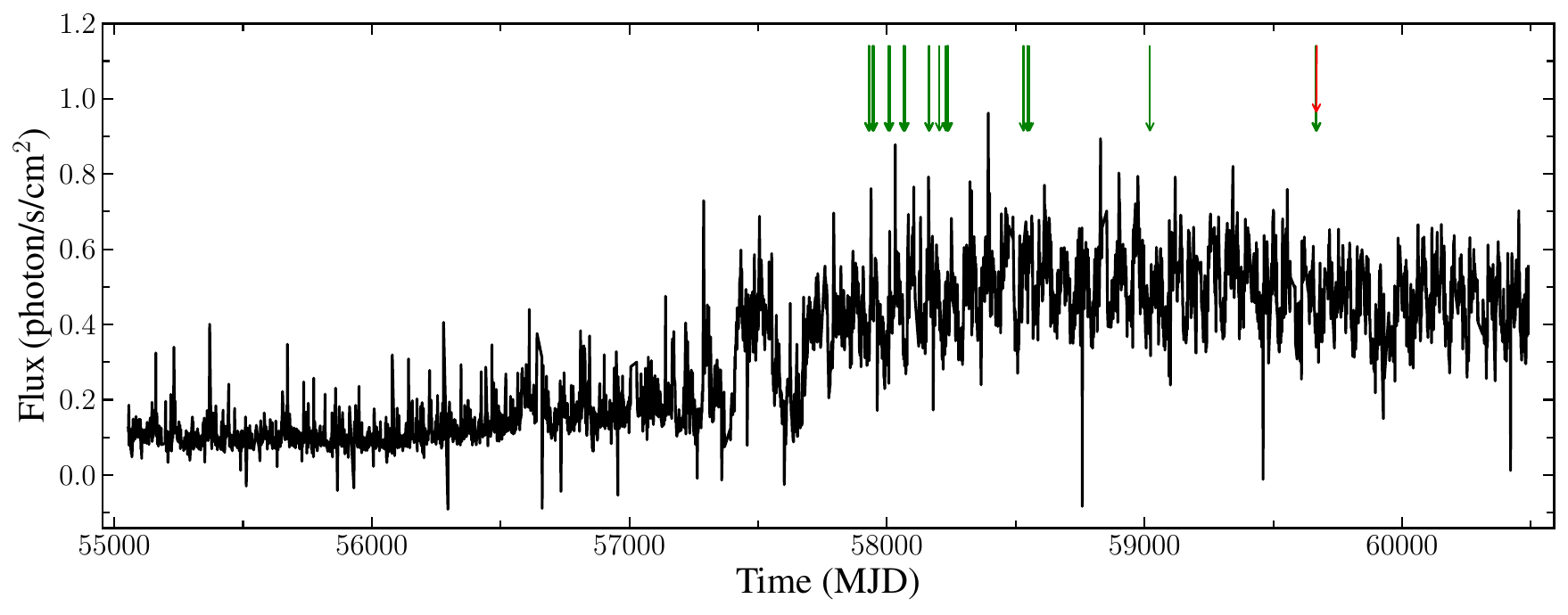}
	\centering
	\caption{The long-term light curve of GS 1826$-$238 observed with \textit{MAXI} in the energy range of 2-20\,keV. The green arrows mark the time of {\it NICER} observations and the red arrow represents the {\it Insight}-HXMT observation. 
	\label{fig:lc}}
\end{figure}

Launched in 2017 June, {\it Neutron star Interior Composition Explorer} \citep[{\it NICER};][]{Gendreau2017} is a payload installed on the International Space Station.
{\it NICER}/X-ray Timing Instrument \citep[XTI;][]{Gendreau2016} consists of 56 X-ray concentrators (XRCs) and silicon drift detector (SDD), providing a large effective area of nearly 1900 cm$^2$ and an energy band of 0.2-12\,keV. 
XTI offers a high time resolution of approximately 100\,ns, which makes it an ideal instrument for timing studies.
In this study, we analysed 41 {\it NICER} observations with an exposure of $>$ 500\,s, spanning from 2017 June to 2022 March (see Figure~\ref{fig:lc}). These data were processed using {\sc heasoft v6.33.2} and {\sc nicerdas} version {\tt 2024-02-09\_V012A}, along with the calibration database (CALDB) version 20240206. 
We filtered the data using {\tt nicerl2}\footnote{\url{https://heasarc.gsfc.nasa.gov/docs/nicer/nicer_analysis.html}} and the latest filter file columns NICERV5.  
We extracted light curves and spectra using the {\tt nicerl3-lc} and {\tt nicerl3-spect} tools respectively.
The background was estimated using the {\tt scorpeon}\footnote{\url{https://heasarc.gsfc.nasa.gov/docs/nicer/analysis_threads/scorpeon-overview/}} model.
The light curves were normalized to an equivalent value of 52 detectors ({\tt detnormtype} parameter was set to {\textquotedblleft}arr52{\textquotedblright} in {\tt nicerl3-lc}).
To extract light curves and spectra for user-defined GTIs, we used the {\tt nimaketime} tool to generate specific GTI files.
%

For each spectrum, we included corresponding systematic errors using the {\tt niphasyserr} task.
Nevertheless, an evident structure was still detected below 1\,keV during the joint {\it NICER}-HXMT fitting, which might be due to an artificial calibration issue \citep{Ludlam2018, Chen2023}.
%
In addition, we used the {\tt ftgrouppha} tool to group spectral channels using an optimal binning scheme \citep{Kaastra2016} and a minimal count of 25.
We used the X-ray spectral fitting package {\sc xspec} version 12.14.0h \citep{Arnaud1996} to perform the spectral analysis. 
{All uncertainties in this paper correspond to a confidence level of 90\%, estimated by running Monte Carlo Markov Chains with a length of 20,000 with a burn-in of 10,000 using the Goodman-Weare algorithm.}

\subsection{{\it Insight}-HXMT}

{\it Insight}-Hard X-ray Modulation Telescope ({\it Insight}-HXMT) is China's first X-ray observatory \citep{Zhang2020}. It mainly consists of three payloads covering the energy range of 1-250\,keV, i.e., Low Energy X-ray telescope for 1-10 keV (\citep[LE;][]{Chen2020}, Medium Energy X-ray telescope for 10-30 keV \citep[ME;][]{Cao2020} and Highe Energy X-ray telescope for 20-250 keV\citep[HE;][]{Liao2020}.
It observed GS 1826$-$238 on 2022 March 28, simultaneously with the {\it NICER} observation (5050310101). 
The data reduction was performed by using {\sc hxmtdas v2.06}\footnote{\url{http://hxmtweb.ihep.ac.cn/software.jhtml}} and the calibration database {\sc caldb v2.07}. The GTIs were generated with the following criteria: (1) the elevation angle  $ > 10^{\circ}$; (2) the time to SAA $ > 300\,$s; (3) the cutoff rigidity value $> 8$\,GeV; (4) the Moon angle $ > 1^{\circ}$; (5) the Sun angle $ > 70^{\circ}$.
The details of screening processes for light curves and spectra can be found in the official user guide.\footnote{\url{http://hxmtweb.ihep.ac.cn/SoftDoc/847.jhtml}}

\section{Results} \label{sec3}
\subsection{Statistics of mHz QPOs } \label{sec3.1}
Since the presence of mHz\,QPOs could be intermittent, we divided each ${\it NICER}$ observation into continuous GTI segments, and only considered those with an exposure of more than 500\,s as an independent sample.

We extracted the light curve in 0.5-10\,keV with a bin size of 1\,s and calculated the Leahy normalized \citep{Leahy1983} power density spectrum (PDS) using the software {\sc stingray} \citep{Huppenkothen2019} (Figure\,\ref{fig:pds}).
We note that the uncertainty in the background is negligible, as it only accounts for less than 1\% of the total count rate.
Clearly, there are PDSs exhibiting a narrow peak in the millihertz band.
In this paper, we define the detection of QPOs if the power exceeds a 
99.99\% significance level. 

In practice, we estimated this significance level using Monte Carlo simulations.
We used the algorithm proposed by \citet{Timmer1995} to produce mimic light curves, and considered the null hypothesis that the power spectrum can be described as a combination of a power-law component (i.e., red noise) and the white noise.
%
%
%
Finally, we identified 37 QPOs among the total 106 samples.
Using the same approach, we also searched for QPOs in ${\it Insight}$-HXMT data, which however only obtained nondetections.
This is likely due to the much smaller effective area of ${\it Insight}$-HXMT/LE and ME.

\begin{figure}[ht!]
	\includegraphics[width=0.4\linewidth]{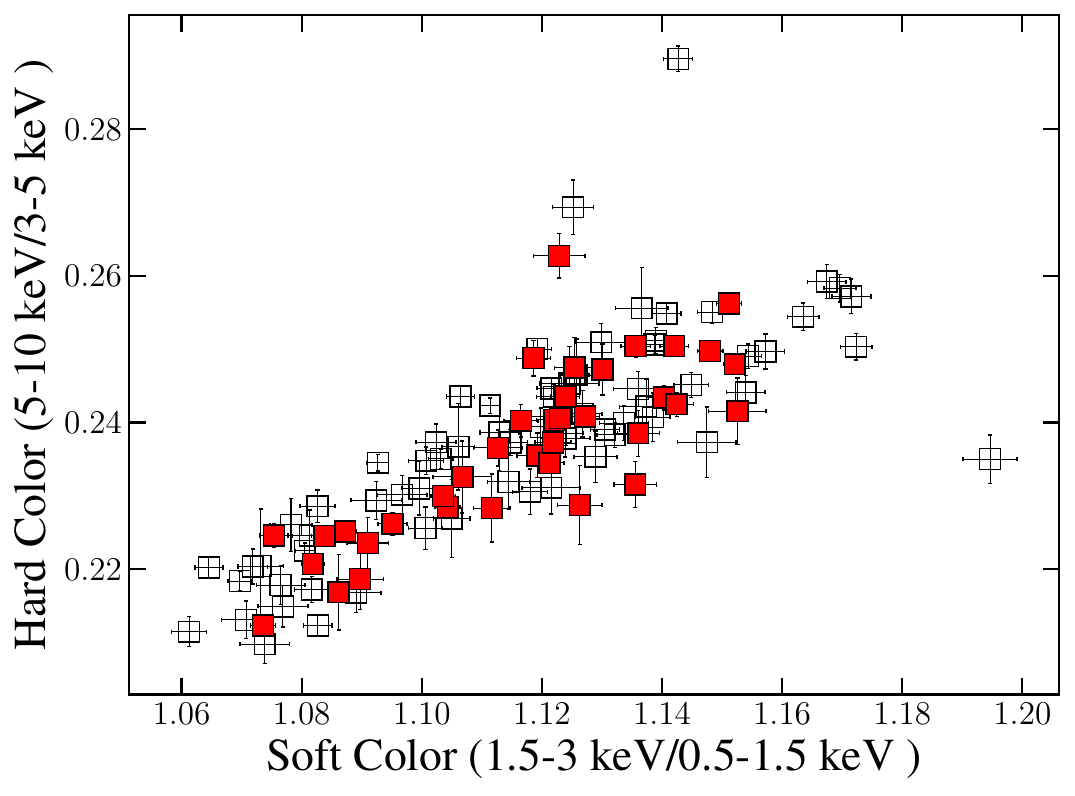}
    \includegraphics[width=0.4\linewidth]{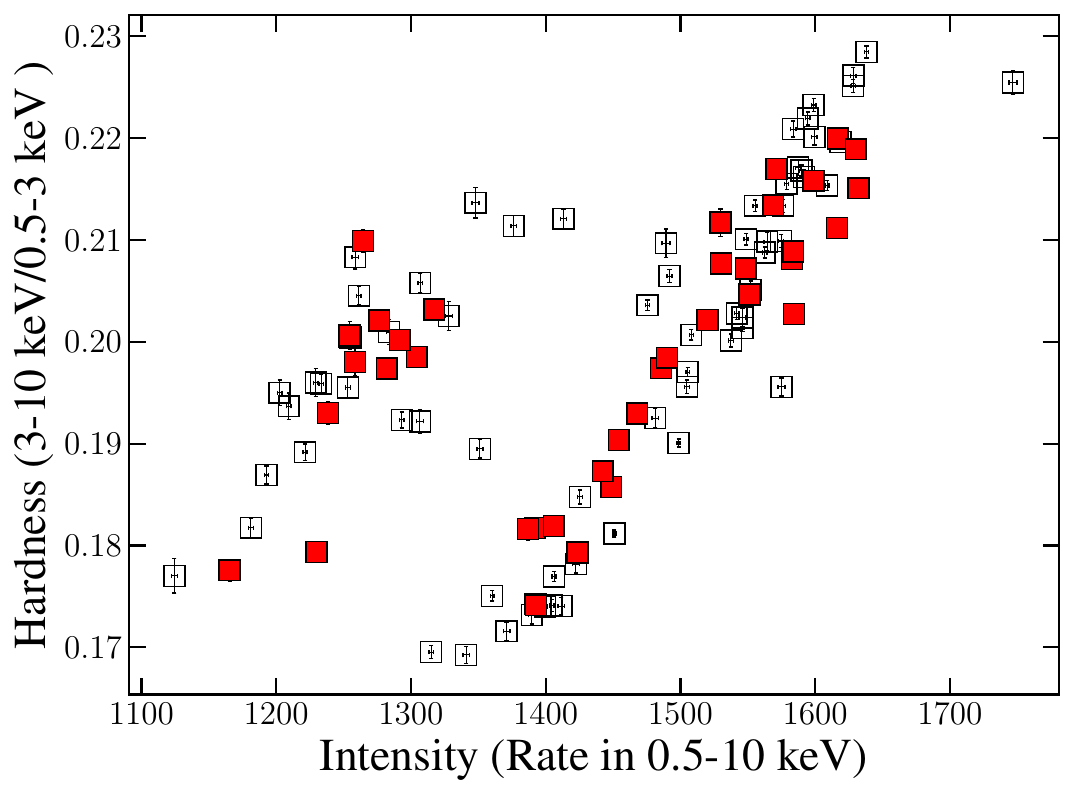}
	\centering
	\caption{Color-color diagram (left) and hardness-intensity diagram (right) of GS 1826$-$238 observed with {\it NICER}.  
    Each point represents a continuous GTI with an exposure time greater than 500\,s.
    The hollow squares represent samples where QPOs were not detected, while the red squares represent observations with QPOs. 
    } 
\label{fig:ccd}
\end{figure}

\begin{figure}[ht!]
	\includegraphics[width=0.75\linewidth]{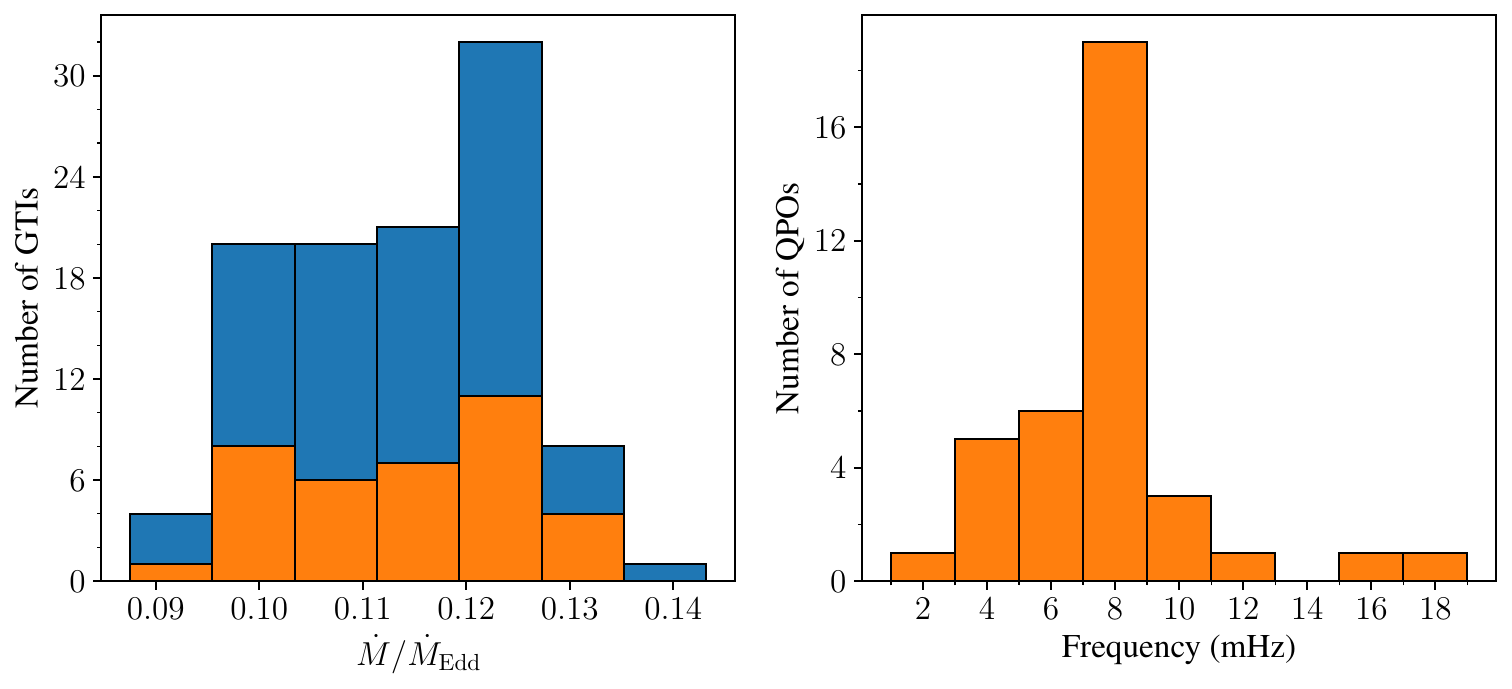}
	\centering
	\caption{Left: accretion rate distributions for all samples (blue) and samples with mHz QPOs (orange). 
    Right: the distribution of mHz QPOs frequencies.
	\label{fig:statisctic}}
\end{figure}

%
To understand the spectral difference between time intervals with and without QPOs, in Figure~\ref{fig:ccd} we compare their positions in the CCD, where the hard (soft) color was defined as the count rate ratio between 5-10\,keV and 3-5\,keV (1.5-3\,keV and 0.5-1.5\,keV).
\citet{Strohmayer2018} suggested that the presence of mHz QPOs might be related to positions in the CCD, preferentially with higher values of hard and soft colors.
However, after including many more observations, this relation is no longer evident.
It appears that there is no significant difference in the CCD between samples with and without QPOs.
In Figure~\ref{fig:ccd}, we also present their positions in the hardness-intensity diagram (HID). 
Generally, the samples can be divided into two branches, each containing observations with and without mHz QPOs.
We used the two-dimensional Kolmogorov-Smirnov test for samples with and without mHz QPOs, assuming a null hypothesis that they are drawn from the same probability distribution in CCD and HID.
The resulting p-values are 0.77 and 0.79 respectively, indicating that these two samples are not significantly different.
After the transition in 2016, GS 1826$–$238 remained in the high-soft state with only minor flux variations. Within a narrow luminosity range, we studied the relation between the presence of mHz QPOs and the accretion rate. The accretion rate was estimated by {the formula $\frac{\dot{M}_i}{\dot{M}_{\rm Edd}} = Rate_{i}*C$, where $C=8 \times 10^{-5}$ is the ratio of the accretion rate obtained from the spectral analysis in Section~\ref{sec3.2} in units of $\dot{M}_{\rm Edd}$ and the count rate, assuming an isotropic radiation of a typical NS ($M_{\rm NS} = 1.4M_{\odot}$).}
As shown in Figure~\ref{fig:statisctic}, the accretion rate ranges from $\sim 0.1\dot{M}_{\rm Edd}$ to $\sim 0.13\dot{M}_{\rm Edd}$.
Although limited by the number of millihertz QPO detections, they appear to occur uniformly over this luminosity range.
The distribution of mHz QPO frequencies, as depicted in Figure~\ref{fig:statisctic} (right panel), is centered around 8\,mHz, with the majority lying below 9 mHz. 
%
%

%

\begin{figure}[ht!]
	\includegraphics[width=0.32\linewidth]{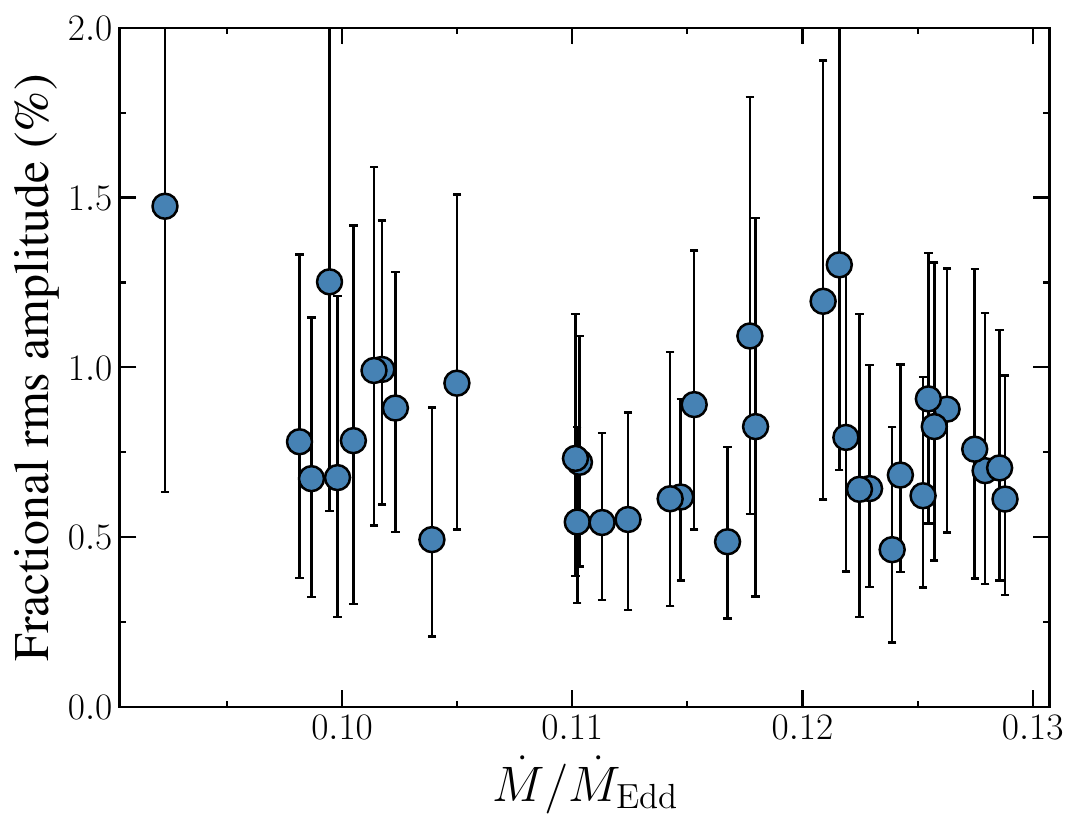}
    \includegraphics[width=0.32\linewidth]{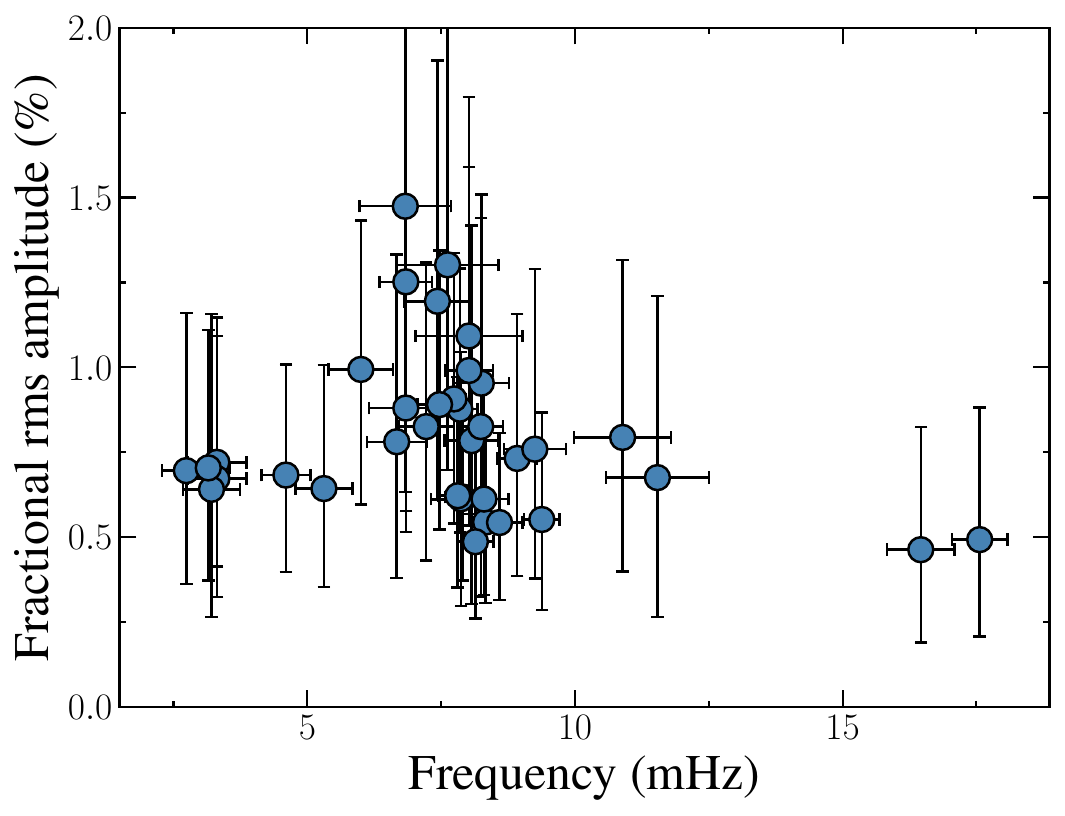}
    \includegraphics[width=0.32\linewidth]{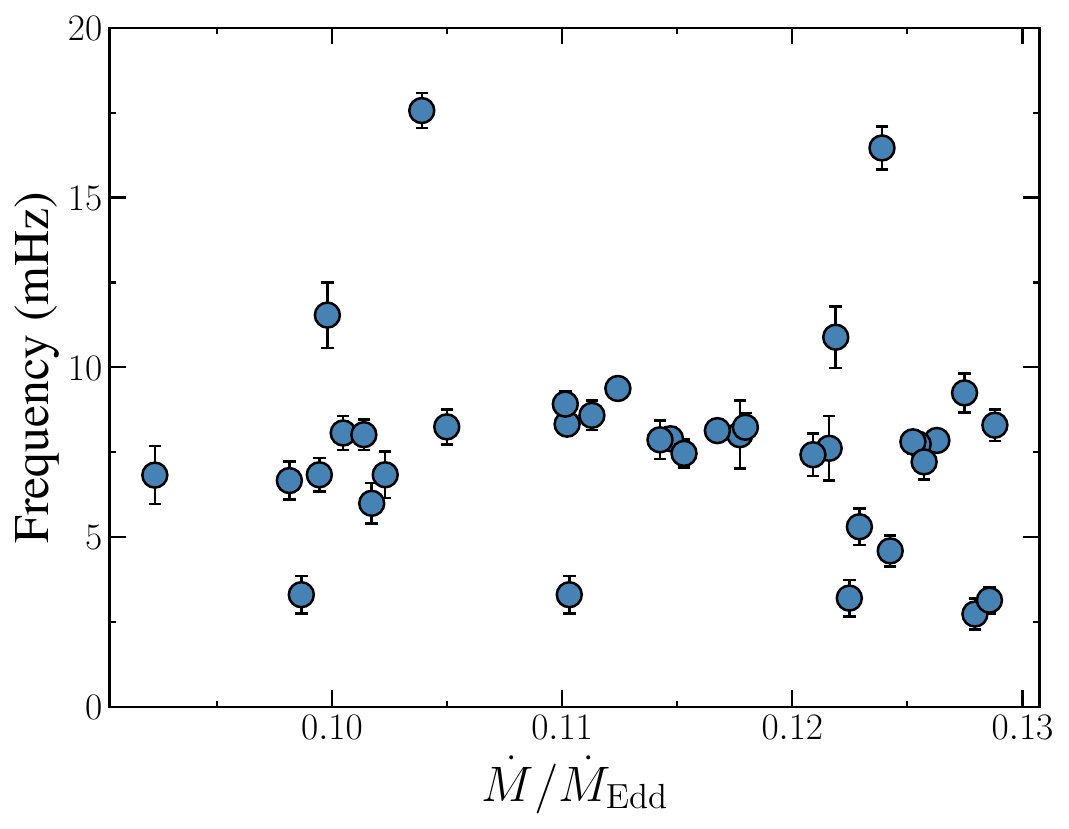}
	\centering
	\caption{Fractional rms amplitude of mHz QPOs vs. accretion rate (left) as well as mHz QPO frequencies (mid). The right panel shows the relation between the QPO frequency and the accretion rate (right). 
	\label{fig:rms}}
\end{figure}

\begin{figure}[ht]
	\includegraphics[width=0.45\linewidth]{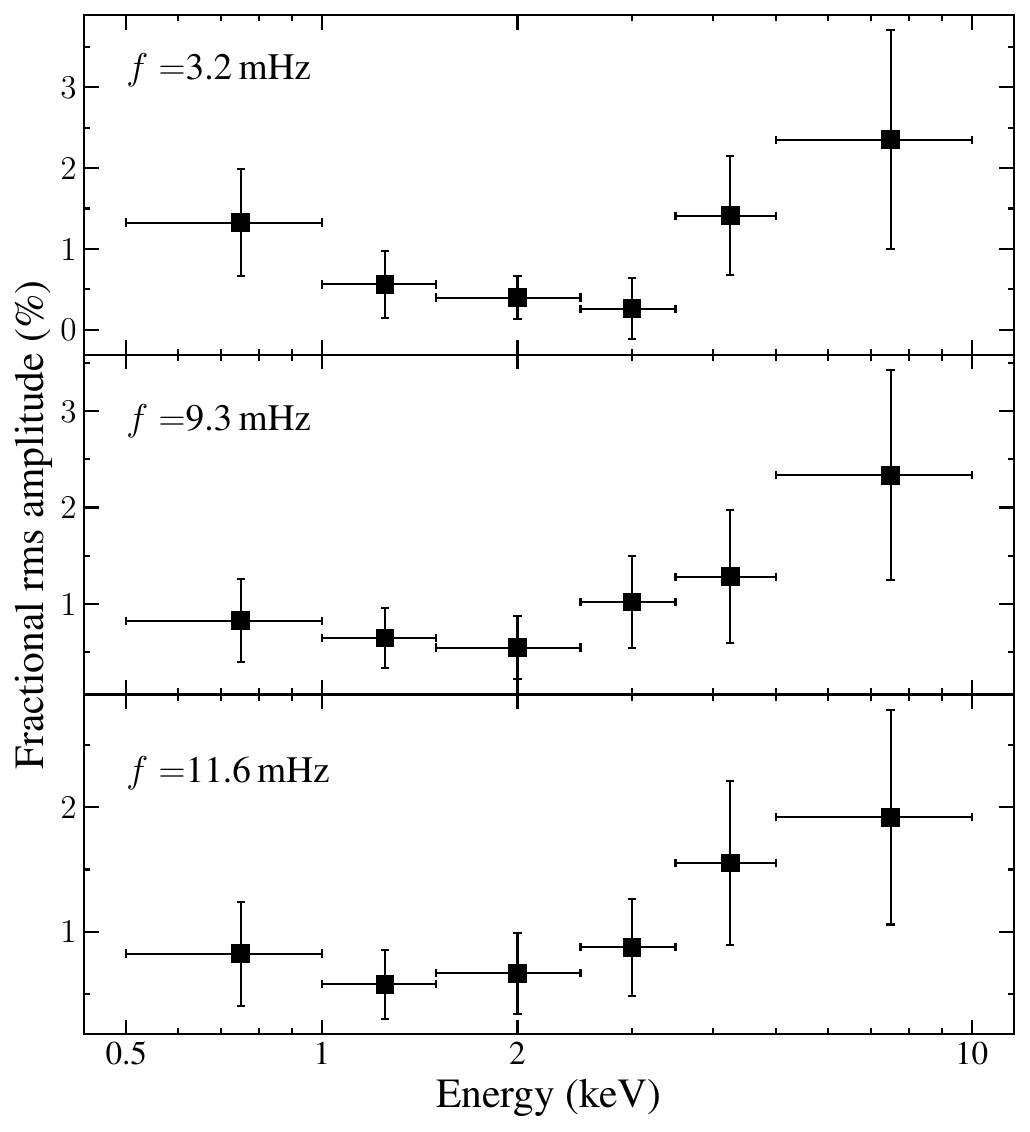}
	\centering
	\caption{Representative fractional rms amplitude of mHz QPO as a function of energy for different QPO frequencies observed with {\it NICER}}. 
	\label{fig:pf_vs_E}
\end{figure}

%
We investigated the evolution of the fractional rms of QPOs with luminosity and energy.
The QPO rms was estimated from power spectra by calculating the square root of the PDS integration around the QPO frequency after subtracting off the white and red noise \citep[see, e.g., equation 3 in][]{Uttley2014}.
The rms uncertainty was obtained from Monte Carlo simulations sampling from observed PDSs.
The rms in the energy range of 0.5-10\,keV turns out to be between $0.5\%$ and $1.5$\%, almost independent of the accretion rate and the QPO frequency (Figure~\ref{fig:rms}).
%
%
We further investigated the dependence of the QPO rms on energy ($E$) and present some representative samples with different QPO frequencies in Figure~\ref{fig:pf_vs_E}.
In all cases, the rms is low for low energies and shows a positive correlation with energy for $E>2$\,keV.
This result is consistent with previous reports by \citet{Strohmayer2018} and in other sources IGR J17480$-$2446 \citep{Linares2012} and 1RXS J180408.9$-$342058 \citep{Tse2021}, but differs from those of 4U 1608$-$52 and 4U 1636$-$536 observed with {\it RXTE} \citep{Revnivtsev2001}.

\subsection{Spectral analysis} \label{sec3.2}
\begin{figure}[ht]
	\includegraphics[width=0.55\linewidth]{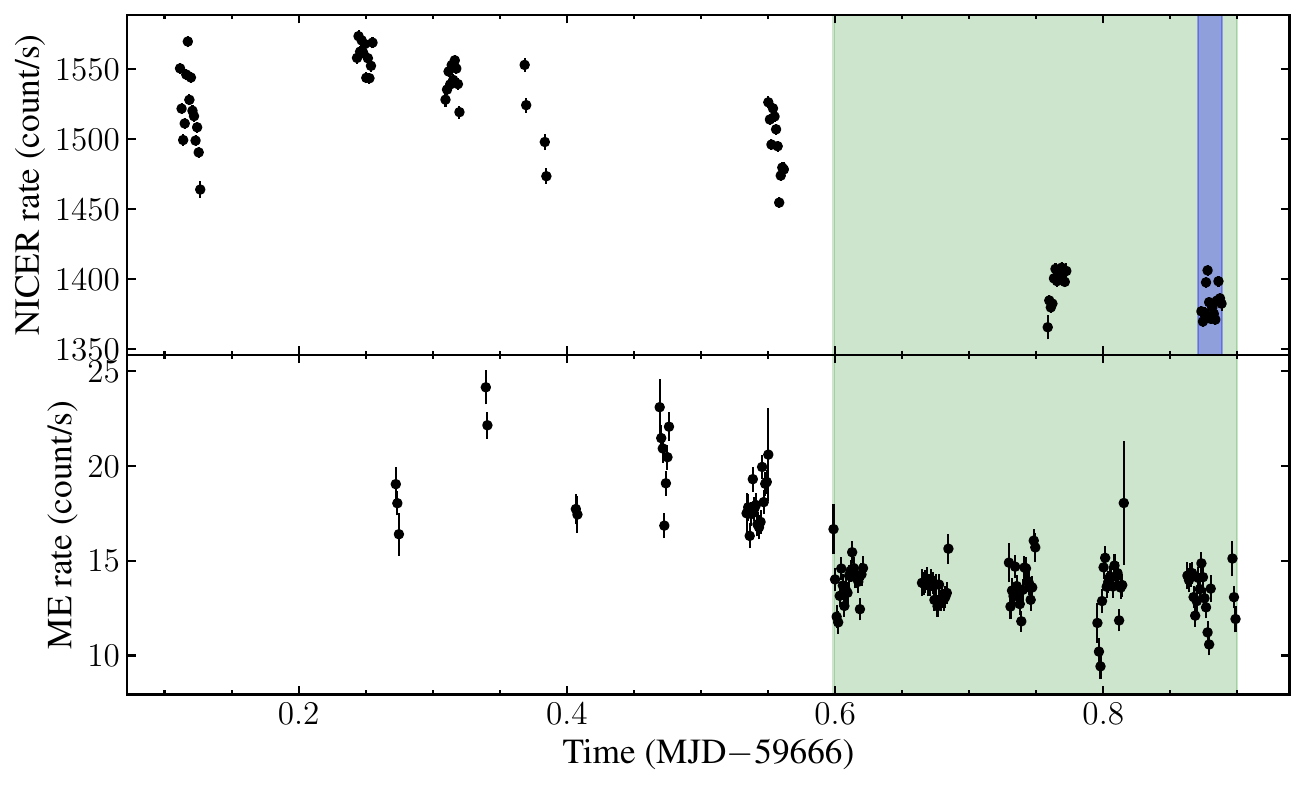}
	\centering
	\caption{Light curves of GS 1826$-$238 on 28 March 2022 observed with {\it NICER} (top) and {\it Insight}-HXMT/ME (bottom), where the blue stripe represents the period with mHz QPOs and the data used for the phase-resolved spectroscopy. The data within the green stripe were used for the averaged spectral analysis. 
	\label{fig:lc_qpo}}
\end{figure}

On 2022 March 28, simultaneous observations of GS 1826$-$238 were carried out with {\it NICER} (5050310101) and {\it Insight}-HXMT (P0404171001). 
We performed joint spectral fitting\footnote{In practice, the good time intervals used are marked in Figure~\ref{fig:lc_qpo}.
} using both observatories to investigate the broadband energy spectroscopy. 
The energy ranges for {\it NICER}, {\it Insight}-HXMT/LE and ME were restricted to 0.7-10\,keV, 2-8\,keV and 10-20\,keV, respectively. 

\begin{figure}[ht!]
	\includegraphics[width=0.5\linewidth]{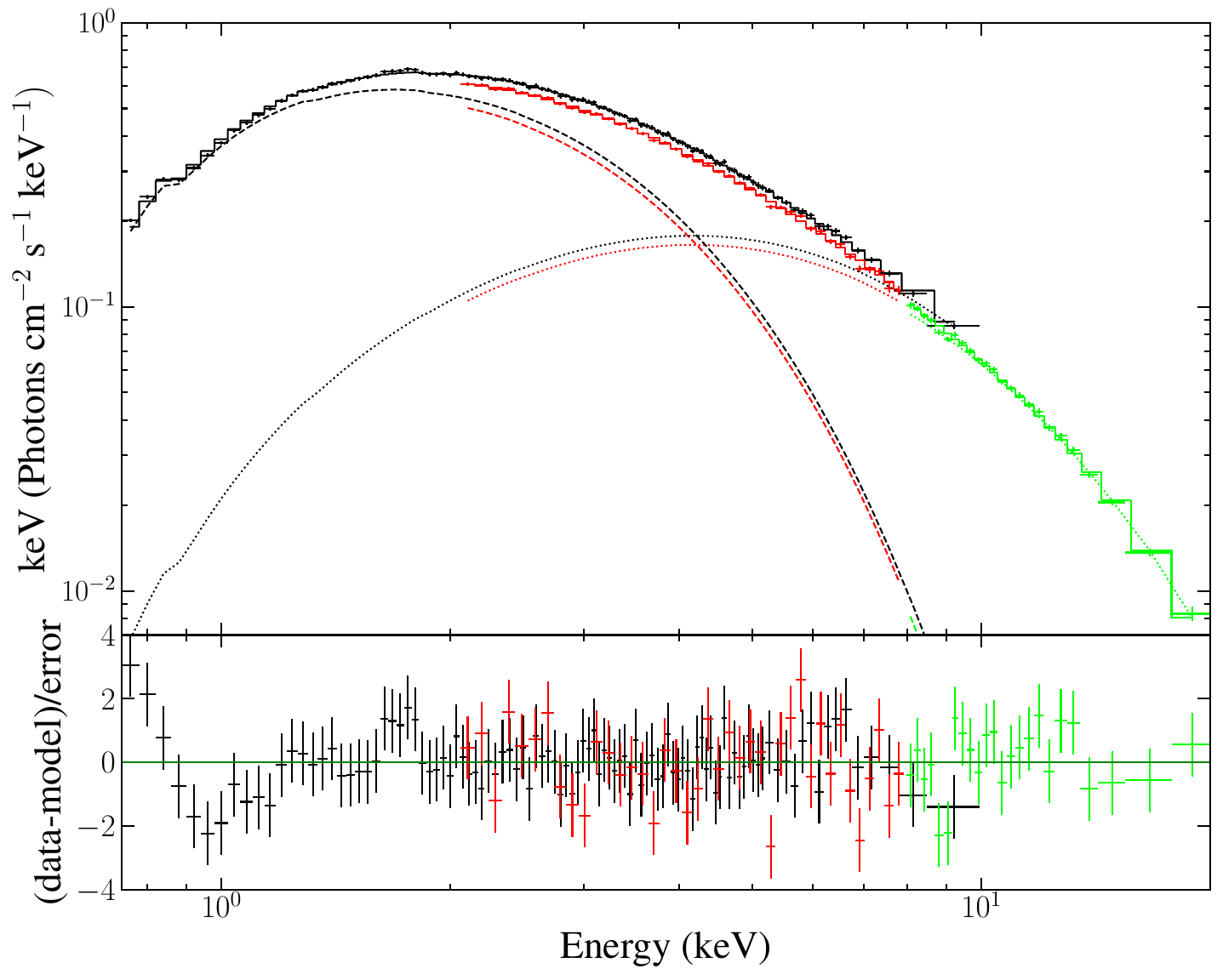}
	\centering
    \caption{
    Unfolded spectra of simultaneous {\it NICER} (black), {\it Insight}-HXMT/LE (red) and ME (green) observations fitted with a spectral model of {\tt const*Tbabs(nthcomp + diskbb)}. 
    The top panel shows the spectra and corresponding model, where the dotted line represents the {\tt nthcomp} component and the dashed line represents the {\tt diskbb} component. 
    The bottom panel shows the residuals. 
	\label{fig:spec}}
\end{figure}

\begin{figure}[ht!]
	\includegraphics[width=0.45\linewidth]{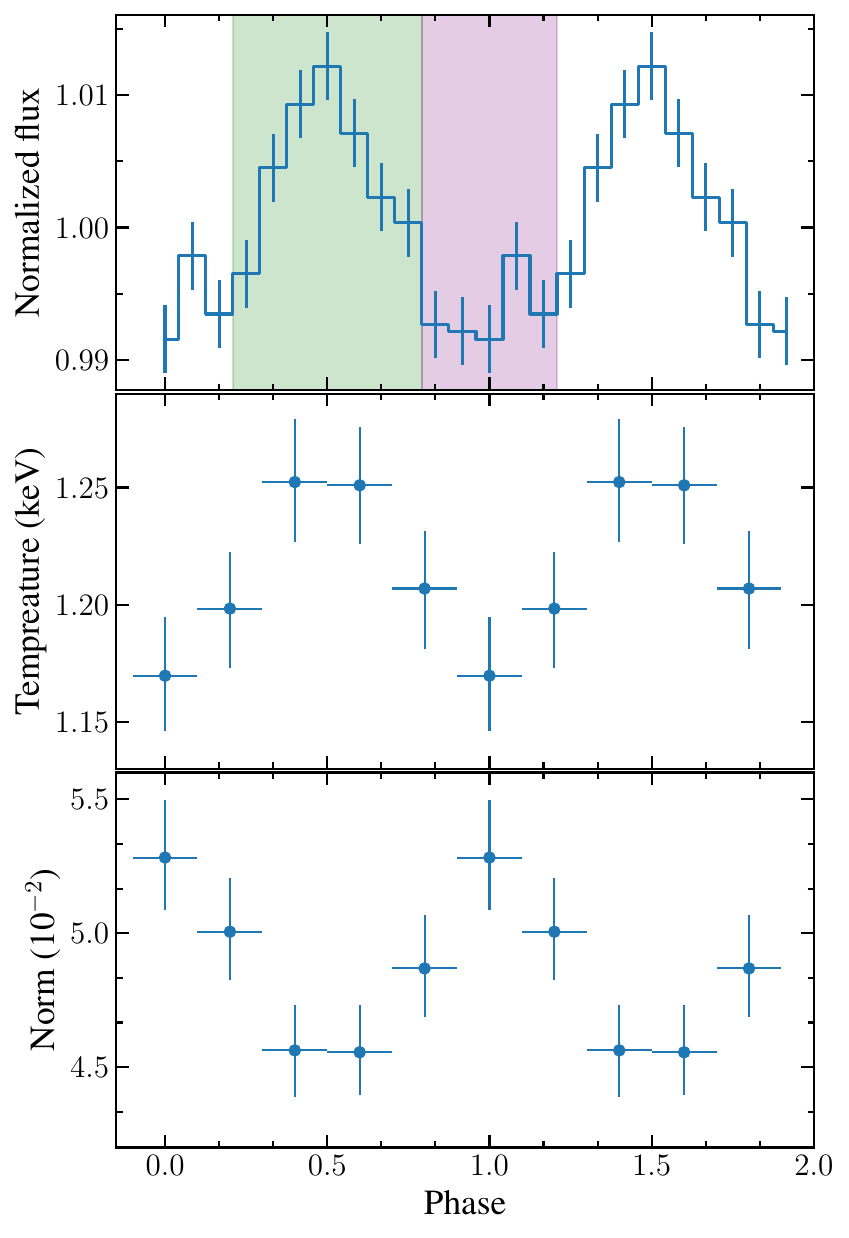}
	\centering 
    \caption{Top panel: the mHz QPO profile of GS 1826$-$238 in the 0.5-10\,keV range discovered by {\it NICER} on 2022 March 28. 
    The green and purple stripes represent the peak and trough phases used in the spectral analysis, respectively. 
    Lower panel: the evolution of the seed photon temperature and the Comptonization normalization.
	\label{fig:para}}
\end{figure}

\begin{figure}[ht!]
	\includegraphics[width=0.47\linewidth]{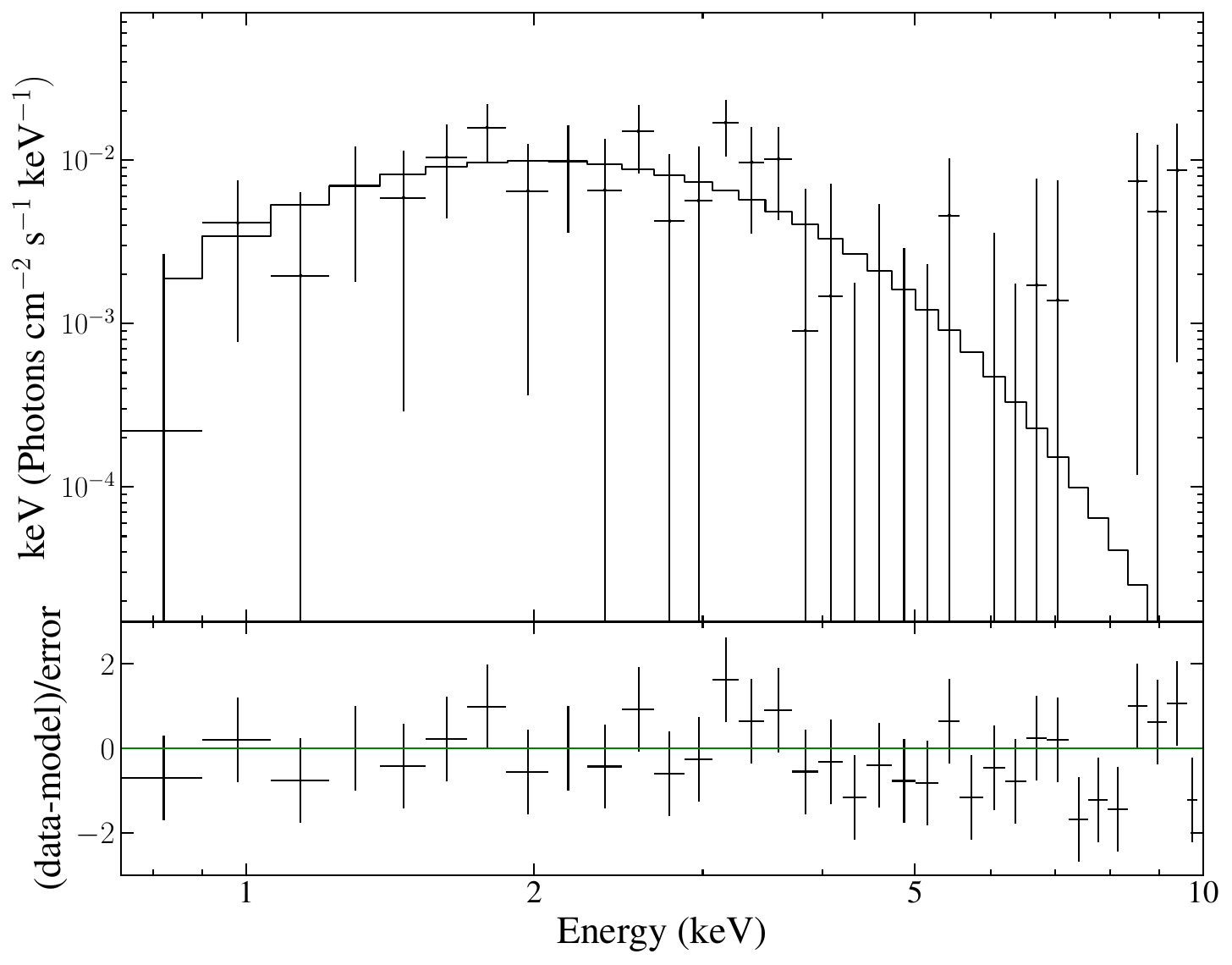}
	\centering
    \caption{Same data as in phase-resolved spectroscopy, but extracting the spectrum during the QPO peak phase and subtracting the spectrum during the QPO trough phase which is considered as the background. 
	\label{fig:spec_bb}}
\end{figure}

According to the shell corona geometry suggested by polarization studies \citep{Capitanio2023}, we adopted a spectral model comprising a disk component ({\tt diskbb}) and a Comptonization component with seed photons originating from the NS surface, i.e., {\tt nthcomp} \citep{Zdziarski1996,Zycki1999} with {\tt inp\_type=0} in {\sc xspec}.
We also included a {\tt Tbabs} model to account for the interstellar absorption \citep{Wilms2000} and a {\tt constant} component for the cross-calibration between instruments.
This model was able to well describe the spectra with a goodness-of-fit $\chi^2\rm/$degrees-of-freedom (dof)=$344.95/380$.
We show the fitting results in Figure~\ref{fig:spec} and Table~\ref{tab:model}.
If assuming a distance of 5.7\,kpc and a viewing angle of $62.5^{\circ}$ \citep{Mescheryakov2011}, the inferred inner radius of the accretion disk is $13.2^{+0.4}_{-0.6}$\rm\,km, which is in agreement with the typical value in NS-LMXBs.
To investigate the spectral properties at different QPO phases, we performed a phase-resolved spectral analysis using {\it NICER} data.
%
%
Here we used the same model with parameters fixed to values obtained from averaged spectra, except for the seed photon temperature ($kT_0$) and the normalization ($norm$) of the Comptonization component. 
{
If the marginally stable burning model is correct \citep{Heger2007}, they correspond to a variable temperature and intensity of seed photons emitted from the NS surface.
The other parameters may describe physical conditions in more extended regions, which are not expected to vary with QPO phases.
}
The results are shown in Table~\ref{tab:model} and Figure~\ref{fig:para}. 
It is evident that both $kT_0$ and $norm$ change over mHz QPO phases and they demonstrate opposite trends.
Typically, the $kT_0$ is positively correlated with the pulse flux, whereas the $norm$ is negatively correlated with the pulse flux.
To study the QPO-modulated spectrum, we extracted the {\it NICER} spectrum during the QPO peak phase and subtracted the spectrum during the QPO trough phase as the background (see Figure~\ref{fig:para}).
This method can provide the information exclusively derived from QPOs, thereby mitigating the spectral degeneracy due to the accretion disk.
Considering mHz QPOs are expected to originate from the marginally stable burning, we employed a blackbody model to describe this spectrum, i.e., {\tt Tbabs*bbodyrad} with the $N_{\rm H}$ fixed at the averaged value.
As shown in Figure~\ref{fig:spec_bb}, this model can provide an acceptable fitting, with goodness-of-fit $\rm \chi^2/dof=95.07/136$. 
The resulting temperature is $0.6^{+0.1}_{-0.1}$\,keV and the normalization is $33^{+39}_{-19}$.
\begin{deluxetable}{cccccccc}
\tablecaption{Best-fitting spectral parameters of the averaged and QPO phase-resolved spectra of GS 1826$-$238. \label{para}}
\tabletypesize{\scriptsize}
\tablewidth{400pt}
\tablenum{1}
\renewcommand\arraystretch{1.4}
\tablehead{
\colhead{Model} & \colhead{Parameter} & \colhead{Average} & 
\colhead{Phase 1} & \colhead{Phase 2} & \colhead{Phase 3} & \colhead{Phase 4} & \colhead{Phase 5} 
} 
\startdata
CONSTANT & $C_{\rm NICER}$ & 1 (fixed)  & fixed & fixed & fixed & fixed & fixed \\
     & $C_{\rm LE}$ &  0.929$^{+0.005}_{-0.004}$  & fixed & fixed & fixed & fixed & fixed \\
     & $C_{\rm ME}$ &  0.882$^{+0.026}_{-0.020}$  & fixed & fixed & fixed & fixed & fixed \\
TBABS & $N_{\rm H}\,(10^{22}\rm cm^{-2})$ & $0.48^{+0.01}_{-0.01}$ & fixed & fixed & fixed & fixed & fixed \\
NTHCOMP & $\Gamma$ & $2.4^{+0.5}_{-0.3}$ & fixed & fixed & fixed & fixed & fixed \\
    & $kT_{\rm e}\,(\rm keV)$ & $3.1^{+0.31}_{-1.28}$  & fixed & fixed & fixed & fixed & fixed \\
    & $kT_0\,(\rm keV)$ & $1.22^{+0.17}_{-0.05}$ & $1.20^{+0.02}_{-0.02}$ & $1.17^{+0.02}_{-0.02}$ &  $1.19^{+0.02}_{-0.03}$ & $1.26^{+0.03}_{-0.03}$ & $1.25^{+0.03}_{-0.03}$ \\
    & $norm\,(10^{-2})$ & $4.8^{+0.5}_{-1.1}$ & $5.0^{+0.2}_{-0.2}$ & $5.2^{+0.2}_{-0.2}$  & $5.3^{+0.2}_{-0.2}$ & $4.5^{+0.2}_{-0.2}$ & $4.7^{+0.2}_{-0.2}$ \\
DISKBB & $kT\,(\rm keV)$  & $0.94^{+0.03}_{-0.02}$ & fixed & fixed & fixed & fixed     & fixed \\
    & $R_{\rm in}\rm\,(km)^{\rm *}$ & $13.2^{+0.4}_{-0.6}$ & fixed & fixed & fixed & fixed & fixed \\
\enddata
\tablecomments{*: The disk inner radius $R_{\rm in}$ was estimated by the formula $N_{\rm d}=(R_{\rm in}/D_{10})^2\rm\, cos\,\theta$, where $N_{\rm d}$ is the normalization of the {\tt diskbb} component, $D_{10}$ the distance to the source in units of 10\,kpc and $\theta$ is the viewing angle of the disk.   
\label{tab:model}}
\end{deluxetable}
\section{discussion and conclusion} \label{sec4}
As a follow-up to the first discovery of \citet{Strohmayer2018}, we report new mHz QPOs detected in GS 1826$-$238 using archival {\it NICER} data spanning 2017 June to 2022 March.
Of 106 segments with a duration of $>$ 500\,s, we found 37 QPOs at a significance level of $>99.99\%$.
The QPO frequency ranges from $\sim3$ to $\sim17$\,mHz.
Similar QPO phenomena have been detected in other NS-LXMBs \citep{Revnivtsev2001,Altamirano2008,Strohmayer2011,Lyu2015,Stiele2016,Mancuso2019,Mancuso2023}.
\citet{Lyu2019} reported a symmetric Gaussian distribution ($\sim$3-14\,mHz) in mHz QPOs in 4U 1636$-$53 with a mean frequency of 8.31 mHz. We tested if our data followed the same distribution using Shapiro-Wilk test, leading to a $p$-value of $1\times 10^{-4}$, which indicates that the samples moderately deviated from the Gaussian distribution we assumed. Therefore, although the frequency range is similar, the frequency distribution of QPOs in our study is asymmetric (see right panel of Figure~\ref{fig:statisctic}), in which mHz QPOs are primarily located in the low-frequency part. 
This may be caused by a much smaller data size in our work, or different properties of sources such as chemical composition, NS magnetic field or accretion rate which affect the mHz QPO mechanism.
In our study, GS 1826$-$238 stays in a soft state, which exhibits moderate spectral evolution as evidenced by its trajectory in the CCDs and HIDs.
We found no correlation between the presence of mHz QPOs and the position of the source in the CCD or HID (see Figure~\ref{fig:ccd}).
This result seems to be inconsistent with previous reports in other sources, in which mHz QPOs only appear at a low hard color state or a high count rate state, such as 4U 1636$-$53 \citep{Lyu2015}, 4U 1730$-$22 \citep{Mancuso2023} and EXO 0748$-$676 \citep{Mancuso2019}.
However, we note that this discrepancy might be only due to the narrow luminosity range of GS 1826$-$238 and the lack of significant spectral evolution.
In theory, mHz QPOs are explained by the marginally stable burning \citep{Heger2007}, which is expected to occur within a narrow range of local accretion rates.
Interestingly, over a long time, GS 1826$-$238 maintained a relatively constant luminosity around $\sim0.10 \dot{M}_{\rm Edd}$ and $\sim 0.13\dot{M}_{\rm Edd}$, coincidentally aligning with the range for marginally stable burning.
Further observations are needed to test the behaviors at different luminosities.

GS 1826$-$238 exhibits a low fractional rms amplitude of mHz QPOs between 0.5\% and 1.5\% in the energy band of 0.5-10\,keV.
According to the marginally stable nuclear burning model \citep{Heger2007}, the QPO frequency and amplitude should be related to the accretion rate $\dot{M}$. 
As $\dot{M}$ increases, the QPOs are expected to evolve toward a stable burning mode, leading to a lower amplitude and a higher frequency. 
In IGR J17480$-$2446, \citet{Linares2012} reported an increase in burst rate and decrease in burst amplitude as the flux rises and considered it as evidence for thermonuclear bursts evolving into a mHz QPO. A similar evolution was reported in Aql X-1 \citep{Li2021}.
\citet{Lyu2019} found an anti-correlation between the fractional rms amplitude of the mHz QPOs and the count rate below
5\,keV in 4U 1636$-$53.
However, as illustrated in Figure~\ref{fig:rms}, our study does not reveal a clear correlation between the QPO amplitude/frequency and the accretion rate. 
This suggests that additional physics should be taken into account, such as the  chemical diffusion and the rotation of the NS \citep{Keek2009}.
On the other hand, we note that these evolutions discovered in IGR J17480$-$2446 and 4U 1636$-$53 span a large range of accretion rates compared to GS 1826$-$238 in our work. Therefore, the QPO amplitude and frequency variations would likely be insignificant for such a stable source.
We performed a spectral analysis of GS 1826$-$238 using simultaneous data from {\it NICER} and {\it Insight}-HXMT. The spectra can be described by a combination of a multicolor blackbody from the accretion disk and a Comptonization component.
The inner disk temperature, estimated to be around 0.9\,keV, suggests a softer spectral shape compared to the latter.
Assuming a spherical geometry, the optical depth of the Comptonization component is $\tau \sim 7.6$ using the equation $\Gamma=-\frac{1}{2}+\sqrt{\frac{9}{4}+\frac{1}{\frac{k T_{e}}{m c^{2}} \tau\left(1+\frac{\tau}{3}\right)}}$ \citep{Zdziarski1996}.
This seems to be consistent with a scenario that QPOs are caused by the marginally stable burning  on the NS surface \citep{Heger2007}, and then emitted seed photons interact with the hot plasma in the optical thick boundary layer with a temperature of a few keV. 
If this is correct, it naturally explains the observed positive correlation between the QPO rms and energy as only the Comptonization component is expected to contribute to mHz QPOs.

In the phase-resolved spectroscopy, we found that the QPO phase-resolved flux was consistent with variations in the seed photon temperature, aligning with the model proposed by \citet{Heger2007}, in which an increase in the temperature of the burning region accelerates the fuel consumption, leading to a reduction in thickness that facilitates more rapid cooling via thermal radiation and therefore forming a pulsed peak.
Our results are consistent with the previous report by \citet{Strohmayer2018}, but differ from those observed in 4U 1636$-$536, where mHz QPOs are caused by a variable spatial extent of the emission region instead of variations in the blackbody temperature \citep{Stiele2016}. 
%
%

%
We detected mHz QPOs in the luminosity range of $\sim 0.1\dot{M}_{\rm Edd} - 0.13\dot{M}_{\rm Edd}$ (Figure~\ref{fig:statisctic}), similar to earlier studies in different sources \citep{Revnivtsev2001,Altamirano2008,Mancuso2019,Mancuso2023}.
However, this contradicts the marginally stable burning model, which predicts that mHz QPOs should occur near the Eddington accretion rate \citep[e.g.,][]{Heger2007}. 
This discrepancy could be explained if mHz QPOs are generated on a specific region of the NS surface, such as near the equator, where the local accretion rate is higher \citep{Heger2007}.
We studied the QPO-modulated spectrum (i.e., the QPO peak spectrum subtracting off the QPO trough spectrum) and found that it can be fitted by a blackbody model with a 
temperature of 0.6\,keV. This implies that the QPO radiation originates from a part of the neutron star surface with a higher local temperature.
We note that there are also alternative solutions for the low observed luminosity, such as the turbulent mixing due to the rotational effect 
and higher heat flux from the crust, which should be explored in the future \citep[for details, see][]{Keek2009, Keek2014}.

\section*{acknowledgments}

This work is supported by the National Natural Science Foundation of China under grant Nos. 12173103 and 12261141691.
This work made use of data from the High Energy Astrophysics Science Archive Research Center (HEASARC), provided by NASA's Goddard Space Flight Center, and from the {\it Insight}-HXMT mission, a project funded by the China National Space Administration (CNSA) and the Chinese Academy of Sciences (CAS).

\appendix
\renewcommand\thefigure{\thesection.\arabic{figure}}    
\setcounter{figure}{0}  
\section{PDS}

\begin{figure}[ht]
    \includegraphics[width=0.245\linewidth]{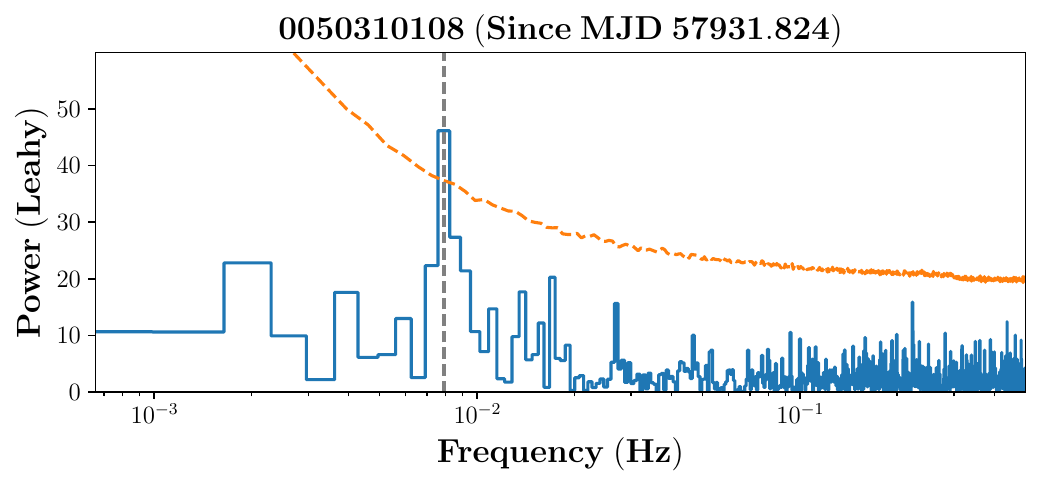}
    \includegraphics[width=0.245\linewidth]{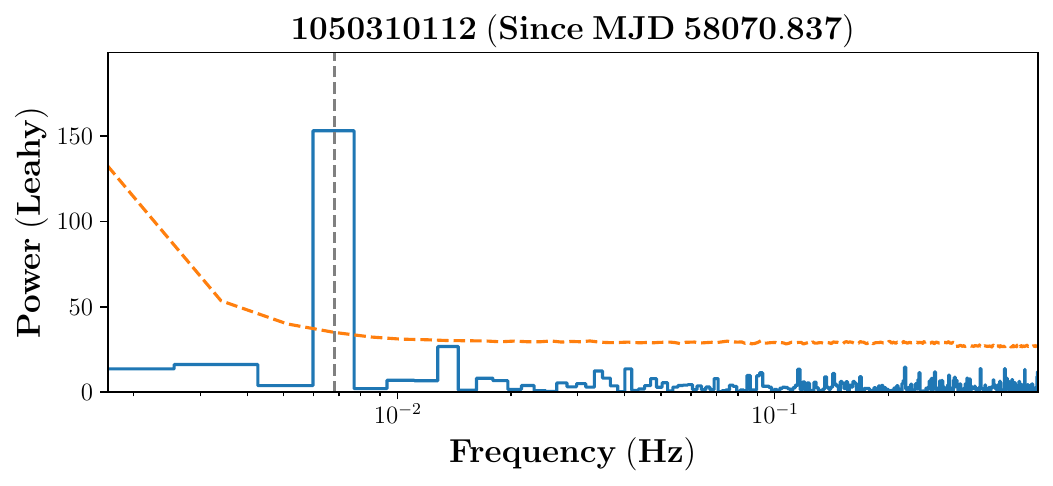}
    \includegraphics[width=0.245\linewidth]{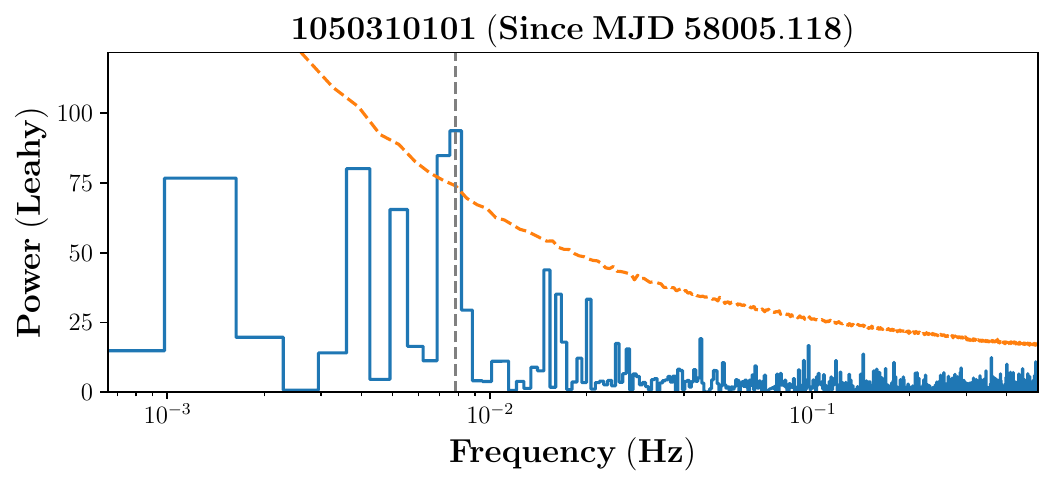}
    \includegraphics[width=0.245\linewidth]{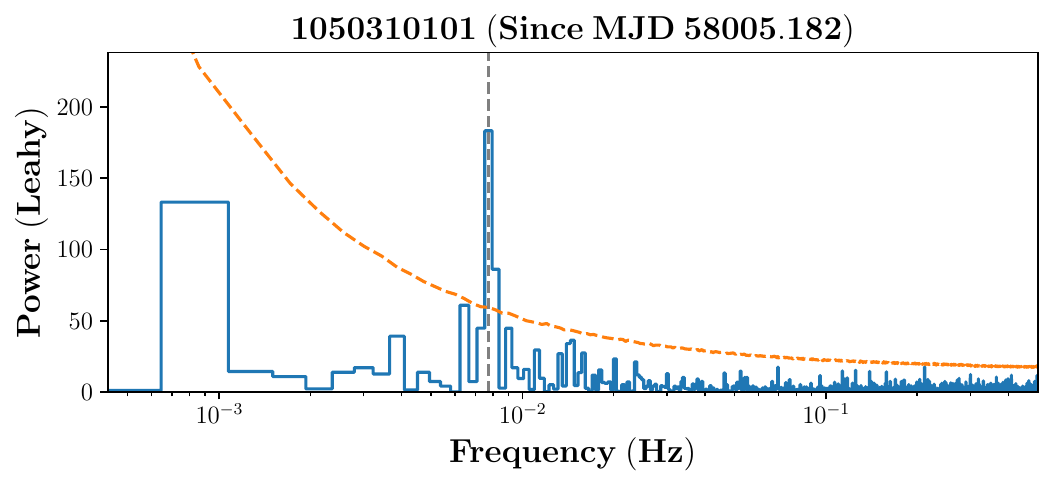}
    \includegraphics[width=0.245\linewidth]{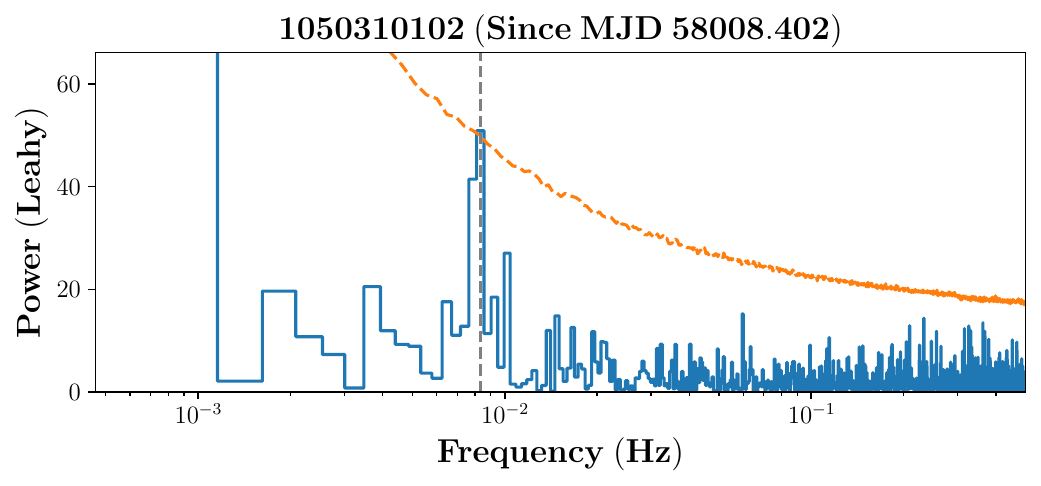}
    \includegraphics[width=0.245\linewidth]{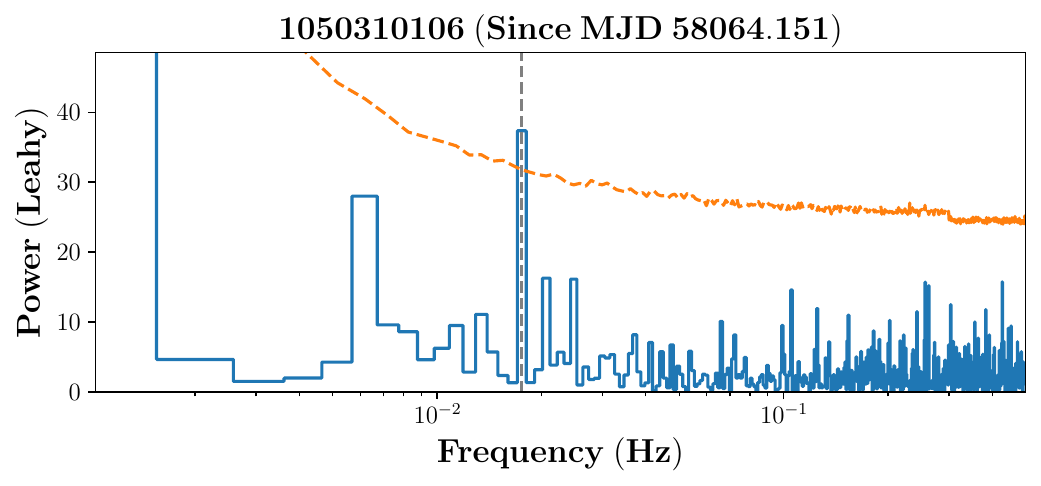}
    \includegraphics[width=0.245\linewidth]{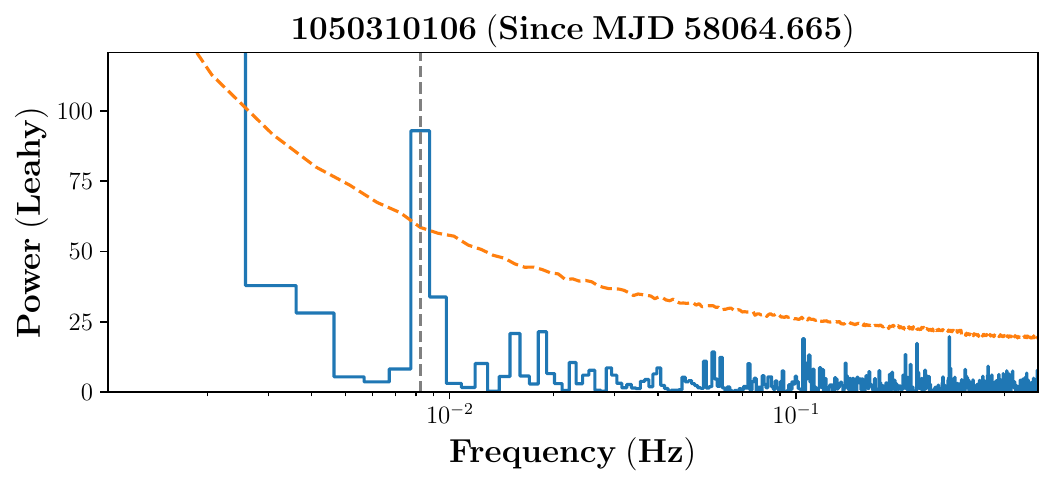}
    \includegraphics[width=0.245\linewidth]{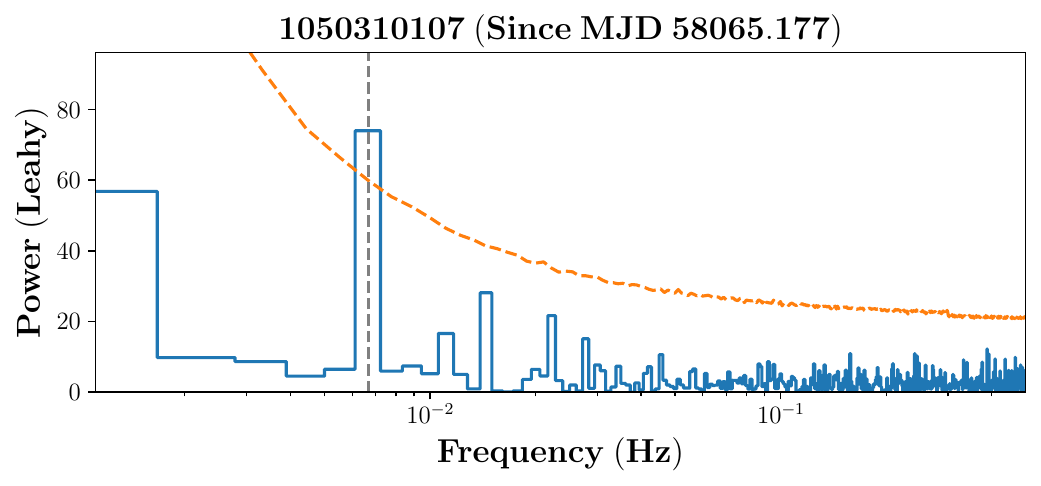}
    \includegraphics[width=0.245\linewidth]{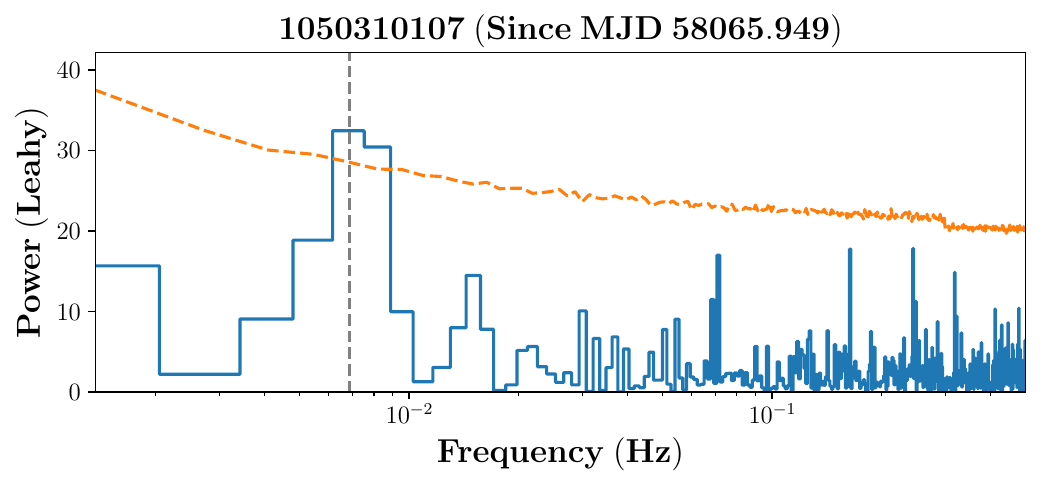}
    \includegraphics[width=0.245\linewidth]{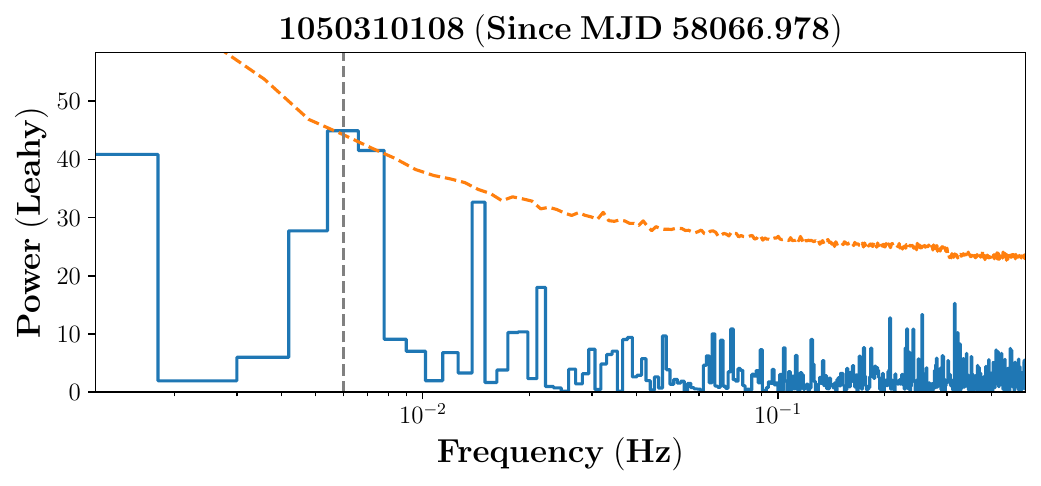}
    \includegraphics[width=0.245\linewidth]{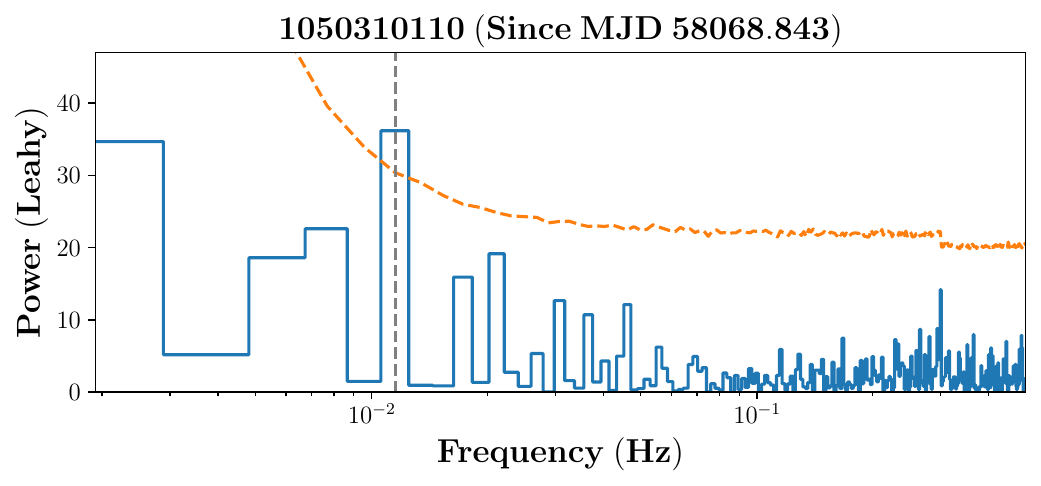}
    \includegraphics[width=0.245\linewidth]{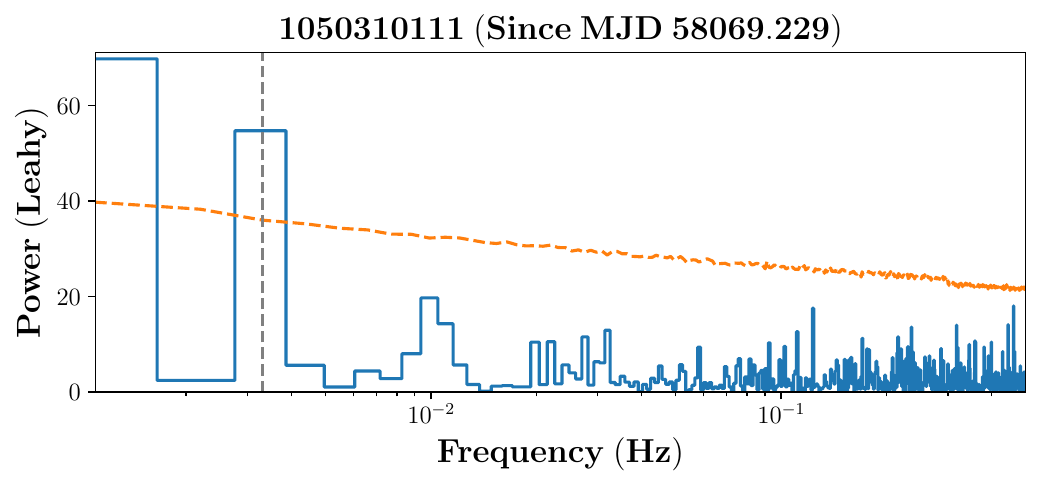}
    \includegraphics[width=0.245\linewidth]{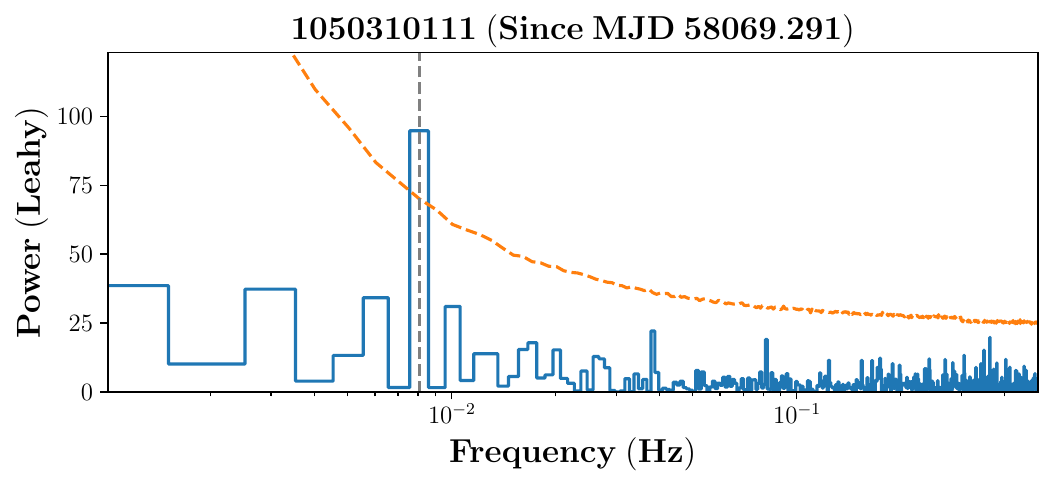}
    \includegraphics[width=0.245\linewidth]{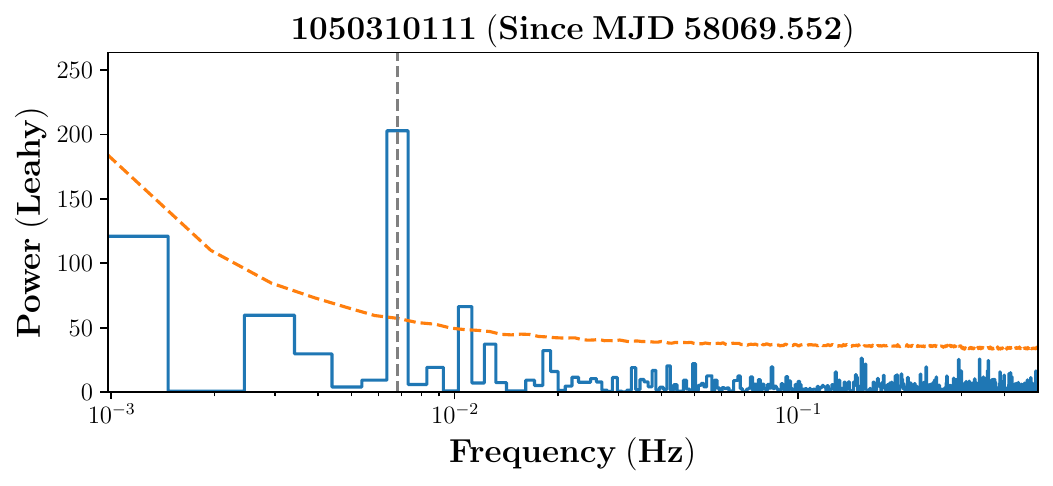}
    \includegraphics[width=0.245\linewidth]{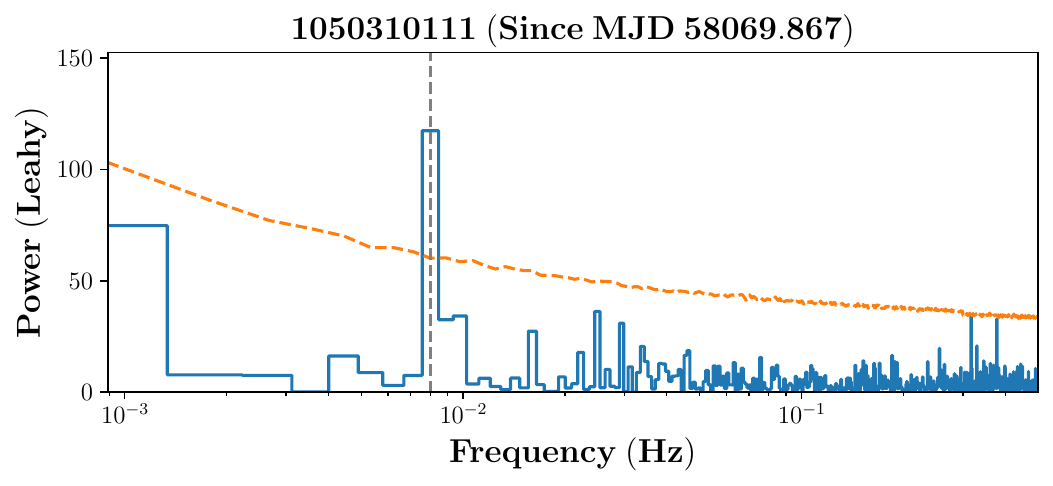}
    \includegraphics[width=0.245\linewidth]{figures/figures/pds_1050310112_3.pdf}
    \includegraphics[width=0.245\linewidth]{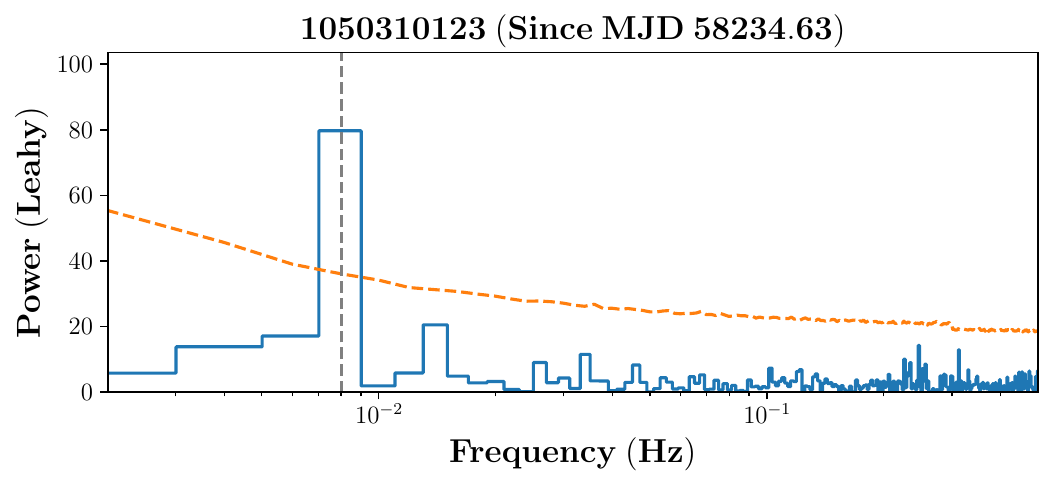}
    \includegraphics[width=0.245\linewidth]{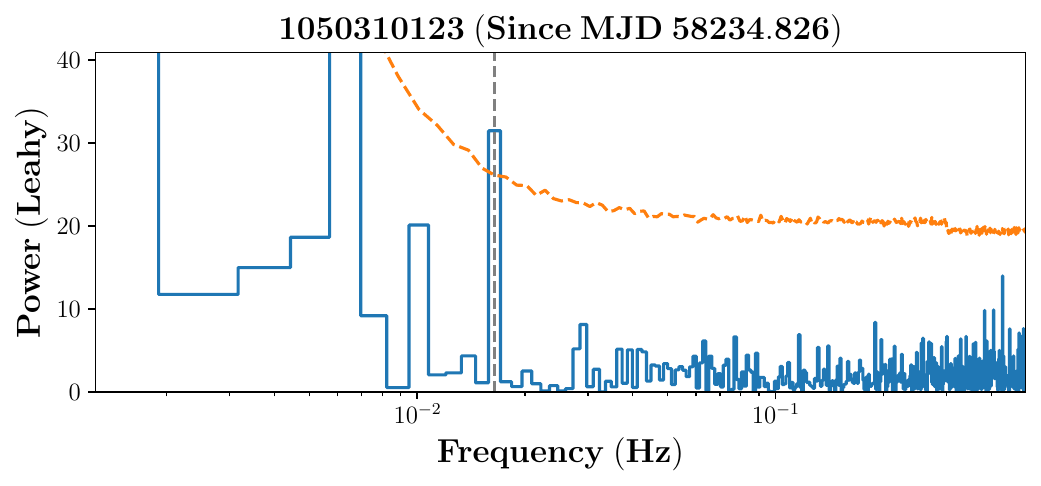}
    \includegraphics[width=0.245\linewidth]{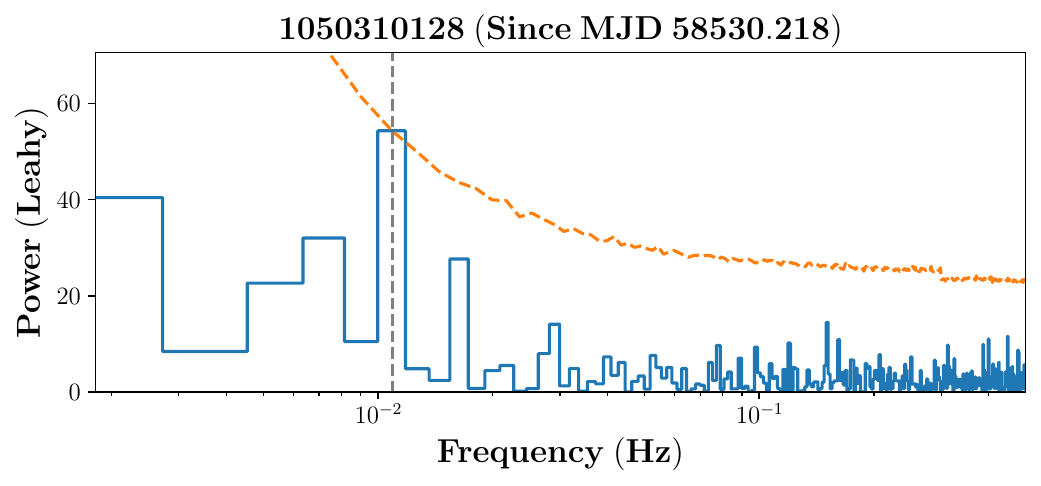}
    \includegraphics[width=0.245\linewidth]{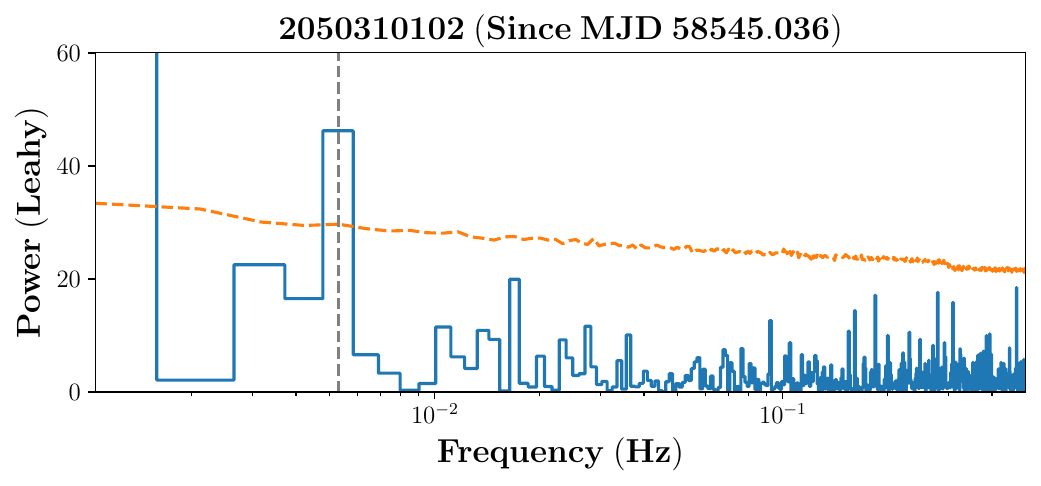}
    \includegraphics[width=0.245\linewidth]{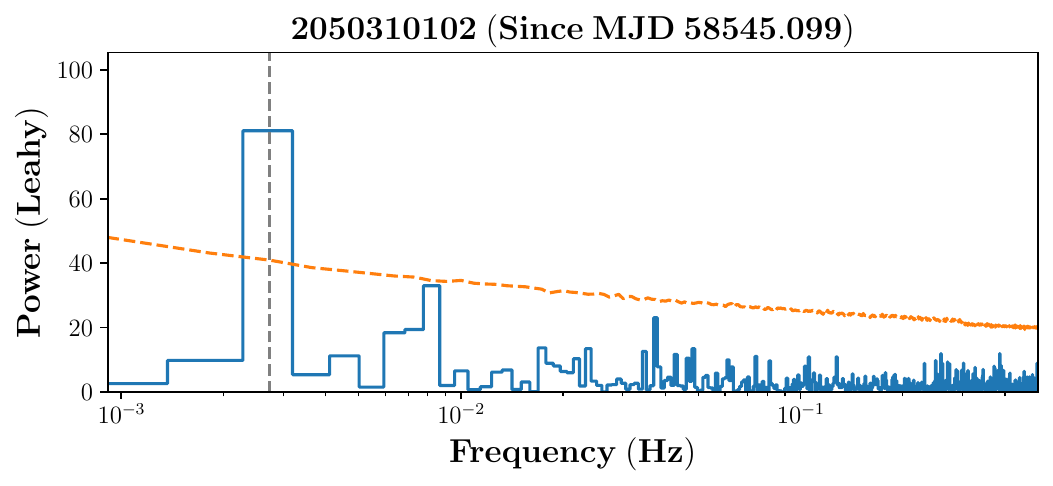}
    \includegraphics[width=0.245\linewidth]{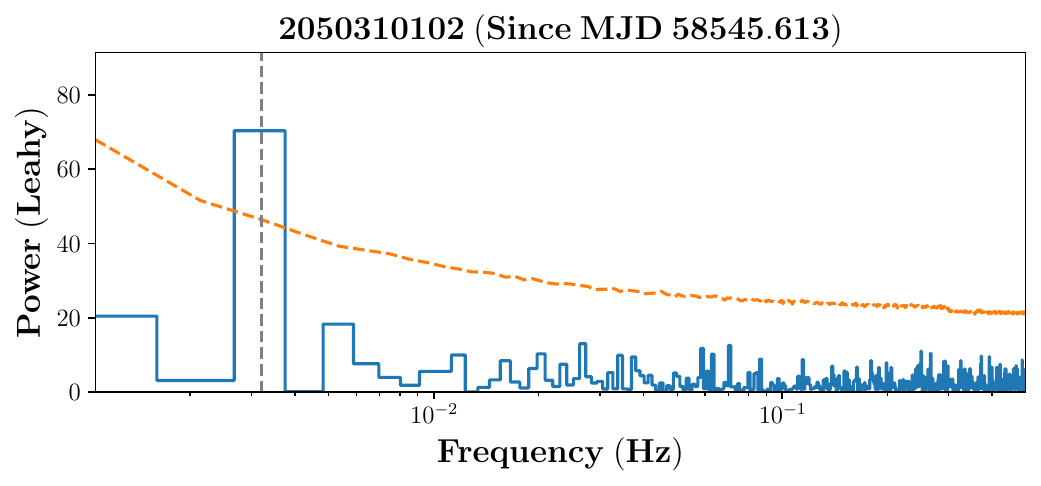}
    \includegraphics[width=0.245\linewidth]{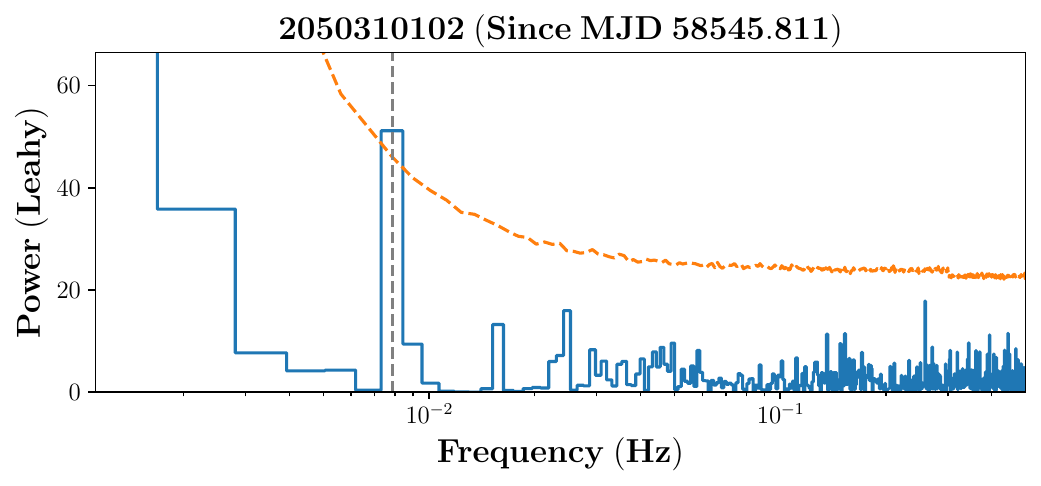}
    \includegraphics[width=0.245\linewidth]{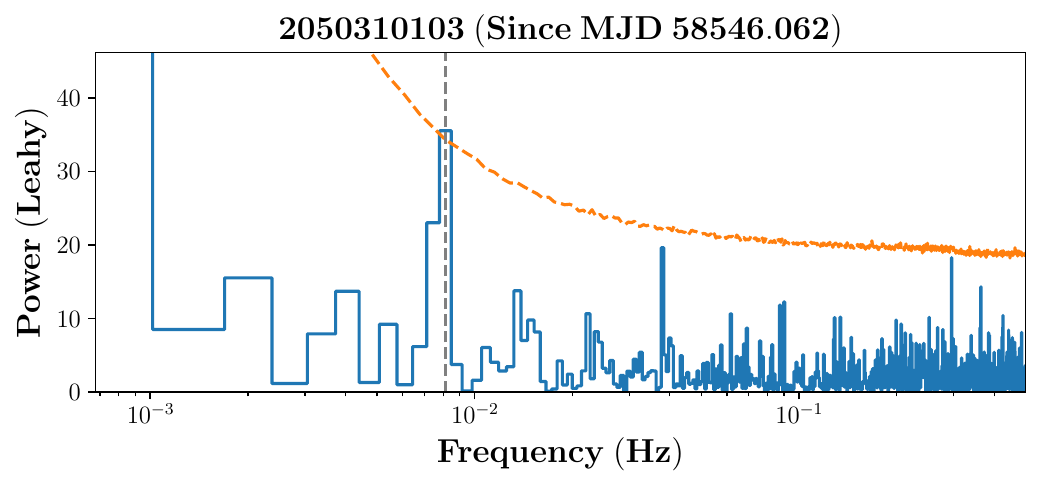}
    \includegraphics[width=0.245\linewidth]{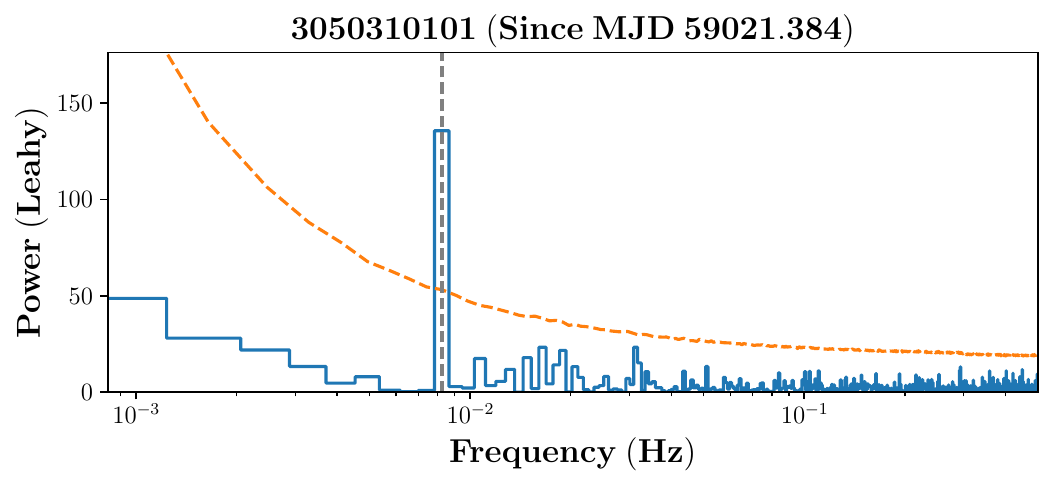}
    \includegraphics[width=0.245\linewidth]{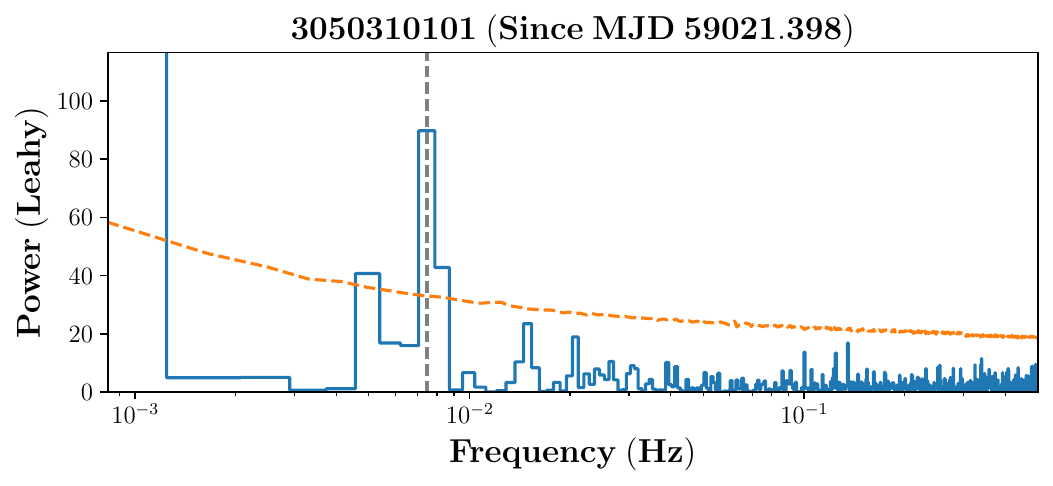}
    \includegraphics[width=0.245\linewidth]{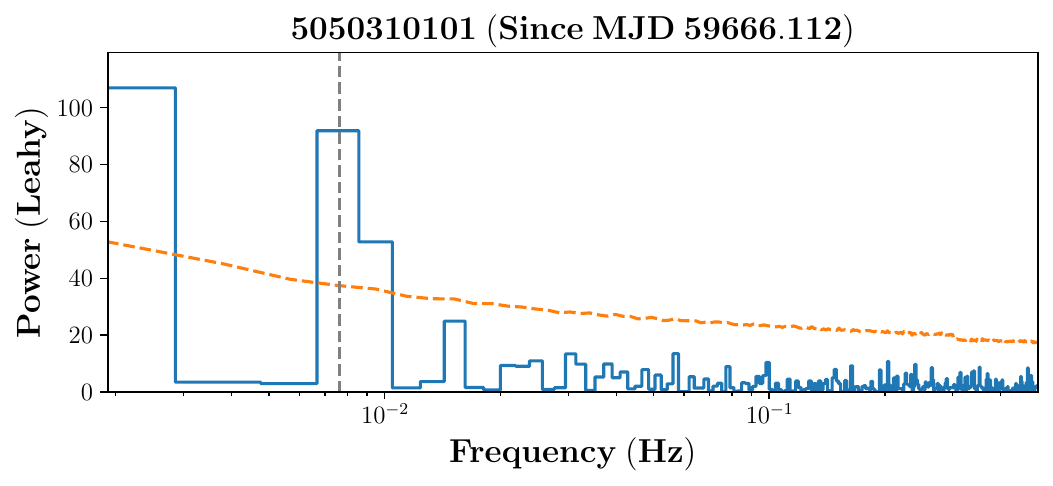}
    \includegraphics[width=0.245\linewidth]{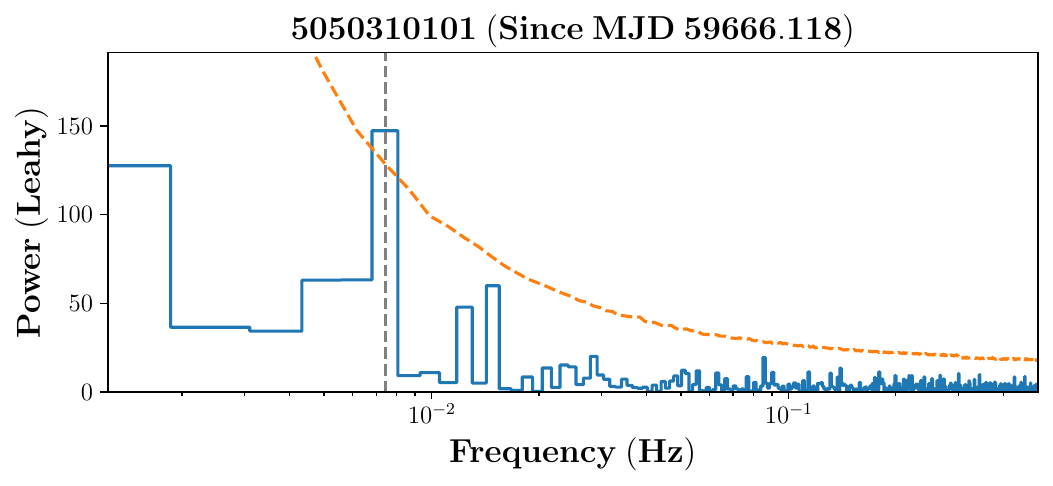}
    \includegraphics[width=0.245\linewidth]{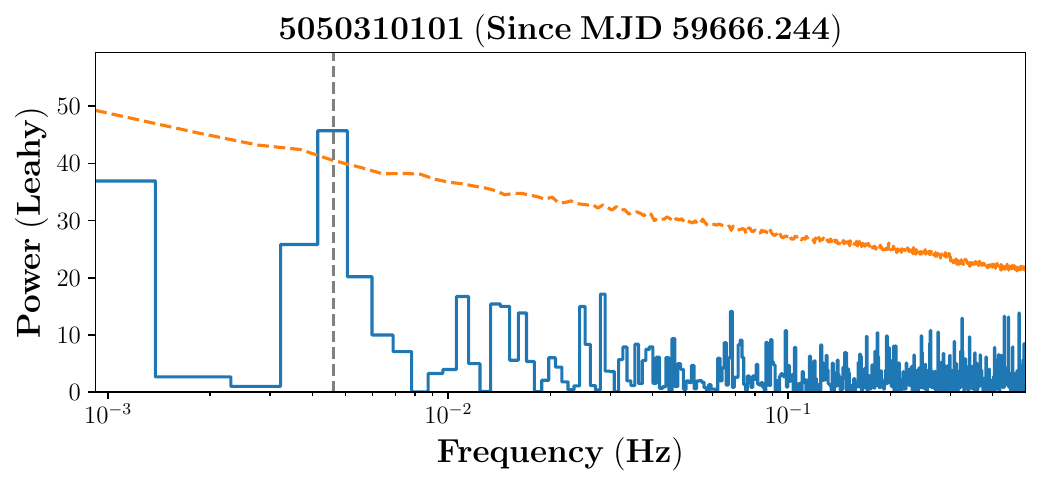}
    \includegraphics[width=0.245\linewidth]{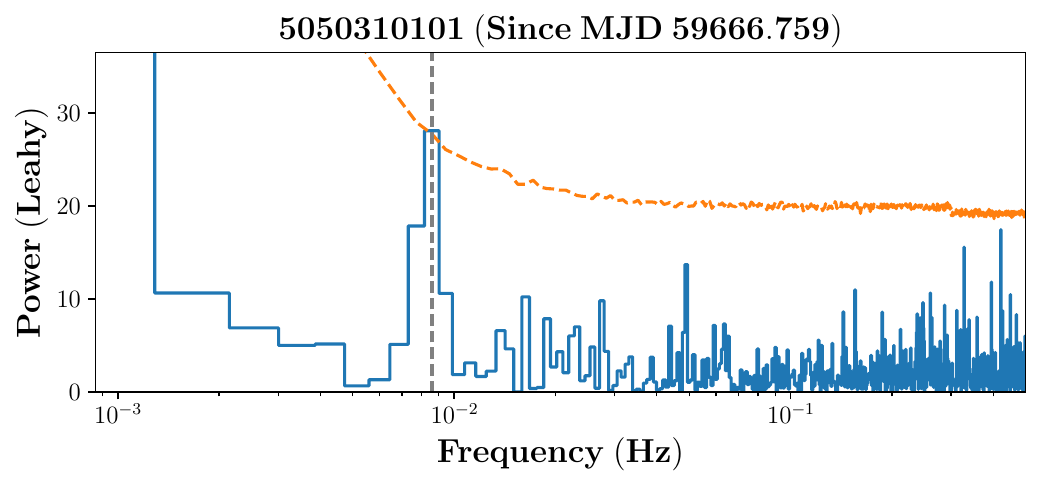}
    \includegraphics[width=0.245\linewidth]{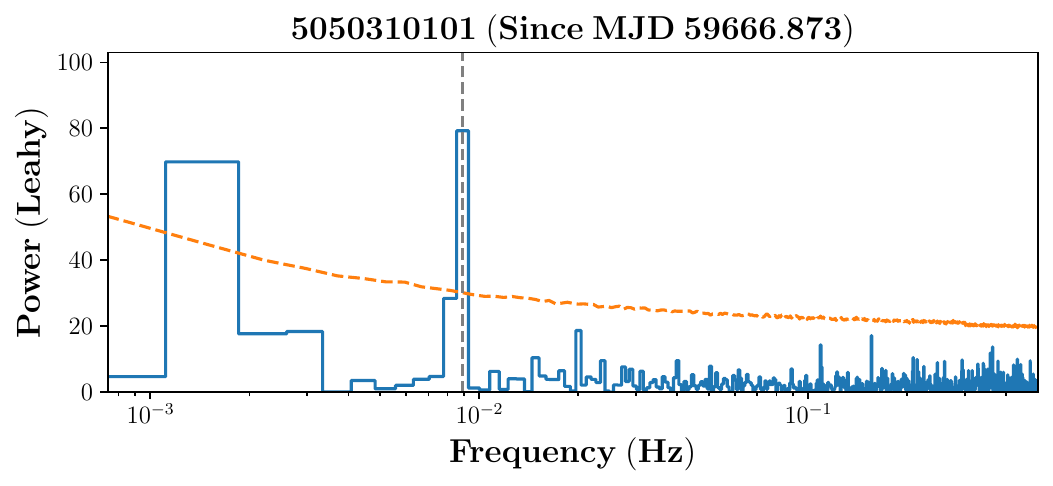}
    \includegraphics[width=0.245\linewidth]{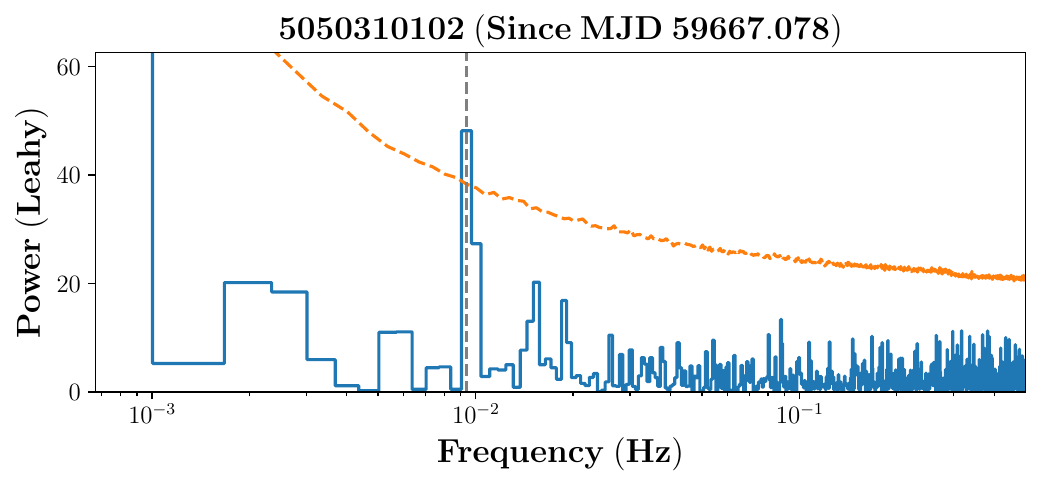}
    \includegraphics[width=0.245\linewidth]{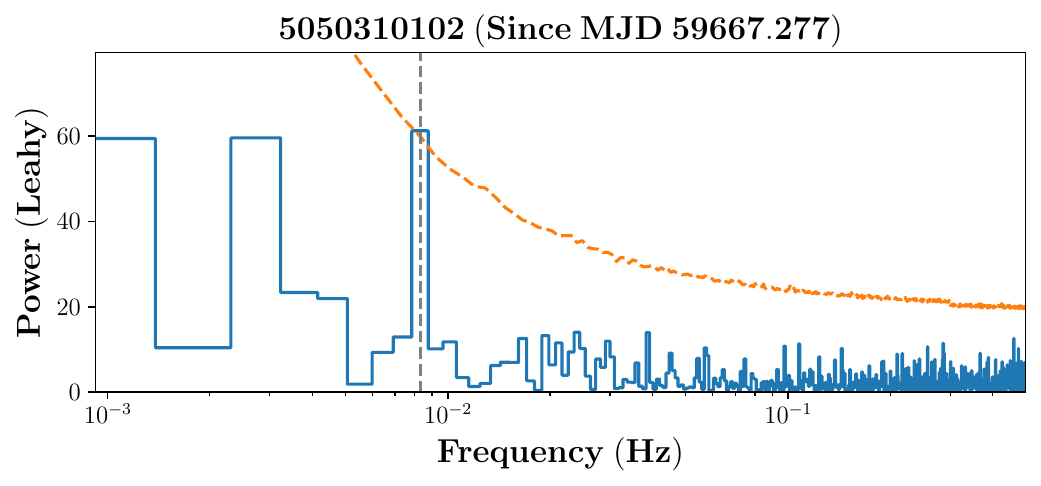}
    \includegraphics[width=0.245\linewidth]{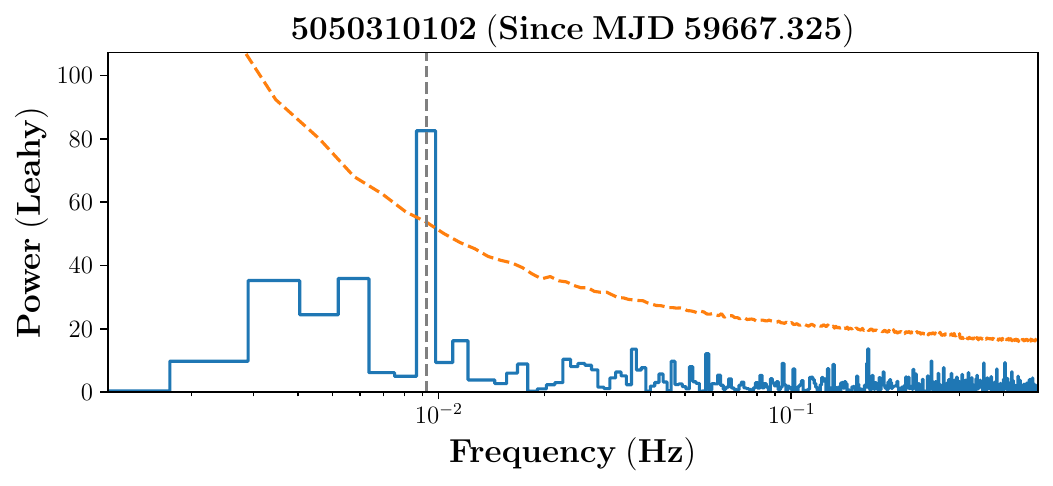}
    \includegraphics[width=0.245\linewidth]{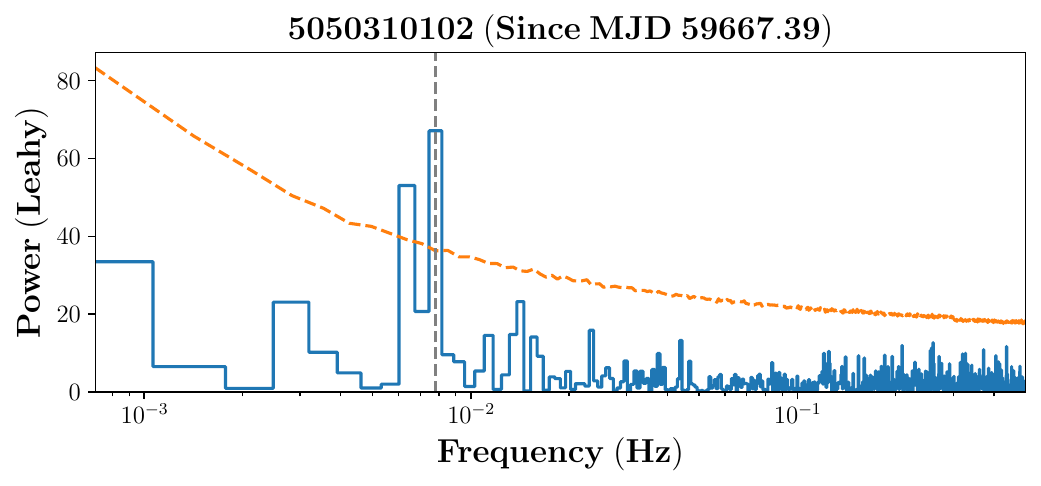}
    \includegraphics[width=0.245\linewidth]{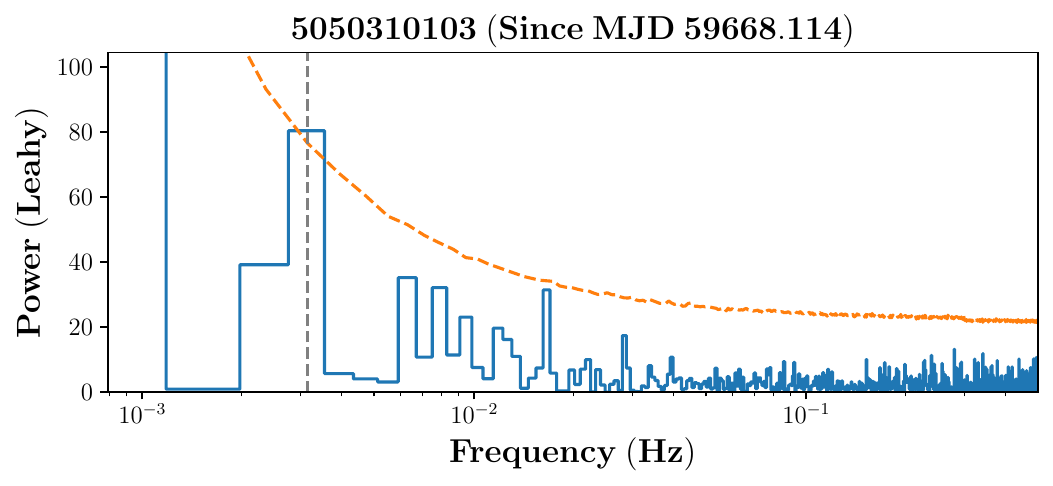}
    \includegraphics[width=0.245\linewidth]{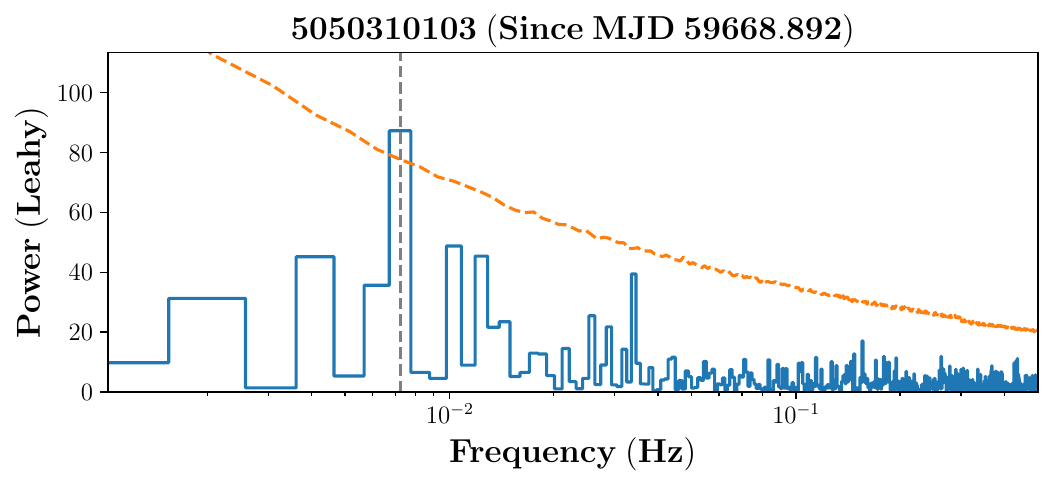}
  \caption{
  PDSs of GTI samples with QPO detections at a significance level $>99.99\%$ in GS 1826$-$238 observed with {\it NICER}. The gray dashed line represents the QPO frequency and the orange dashed line indicates the 99.99\% power threshold.
  }
  \label{fig:pds}
\end{figure}

In Figure~\ref{fig:pds}, we present the PDS for all samples in which mHz QPOs were detected.

\bibliography{main}{}
\bibliographystyle{aasjournal}
\end{document}